\documentclass[journal,twoside,web]{ieeecolor}
\usepackage{generic}
\usepackage{cite}
\usepackage{amsmath,amssymb,amsfonts}
\usepackage{algorithmic}
\usepackage{graphicx}
\usepackage{textcomp}
\def\BibTeX{{\rm B\kern-.05em{\sc i\kern-.025em b}\kern-.08em
    T\kern-.1667em\lower.7ex\hbox{E}\kern-.125emX}}

\begin{document}
\title{Sparse-based Domain Adaptation Network for OCTA Image Super-Resolution Reconstruction}
\author{Huaying Hao, Cong Xu, Dan Zhang, Qifeng Yan, Jiong Zhang*, Yue Liu*, Yitian Zhao


\thanks{ H.~Hao and Y.~Liu are with Beijing Engineering Research Center of Mixed Reality and Advanced Display,School of Optics and Photonics, Beijing Institute of Technology, Beijing, China, 100081.}
\thanks{ H.~Hao, C.~Xu, Q.~Yan, J.~Zhang and Y.~Zhao are with Cixi Institute of Biomedical Engineering, Ningbo Institute of Materials Technology and Engineering, Chinese Academy of Sciences, Ningbo, China.}
\thanks{ D. Zhang is with School of Cyber Science and Engineering, Ningbo University of Technology, Ningbo, China.}
\thanks{\textit{*Corresponding authors: Jiong Zhang (zhangjiong@nimte.ac.cn) and Yue Liu (liuyue@bit.edu.cn)}}}

\maketitle

\begin{abstract}
Retinal Optical Coherence Tomography Angiography (OCTA) with high-resolution is important for the quantification and analysis of retinal vasculature. However, the resolution of OCTA images is inversely proportional to the field of view at the same sampling frequency, which is not conducive to clinicians for analyzing larger vascular areas. 
In this paper, we propose a novel \textbf{S}parse-based domain \textbf{A}daptation \textbf{S}uper-\textbf{R}esolution network (SASR) for the reconstruction of realistic $6\times6~mm^2$/low-resolution (LR) OCTA images to high-resolution (HR) representations. 
To be more specific, we first perform a simple degradation of the $3\times3~mm^2$/high-resolution (HR) image to obtain the synthetic LR image. An efficient registration method is then employed to register the synthetic LR with its corresponding $3\times3~mm^2$ image region within the $6\times6~mm^2$ image to obtain the cropped realistic LR image. We then propose a multi-level super-resolution model for the fully-supervised reconstruction of the synthetic data, guiding the reconstruction of the realistic LR images through a generative-adversarial strategy that allows the synthetic and realistic LR images to be unified in the feature domain. Finally, a novel sparse edge-aware loss is designed to dynamically optimize the vessel edge structure. Extensive experiments on two OCTA sets have shown that our method performs better than state-of-the-art super-resolution reconstruction methods. In addition, we have investigated the performance of the reconstruction results on retina structure segmentations, which further validate the effectiveness of our approach.

\end{abstract}


\section{Introduction}
\label{sec:introduction}
\begin{figure}[t]
\centering{
\includegraphics[width=9cm]{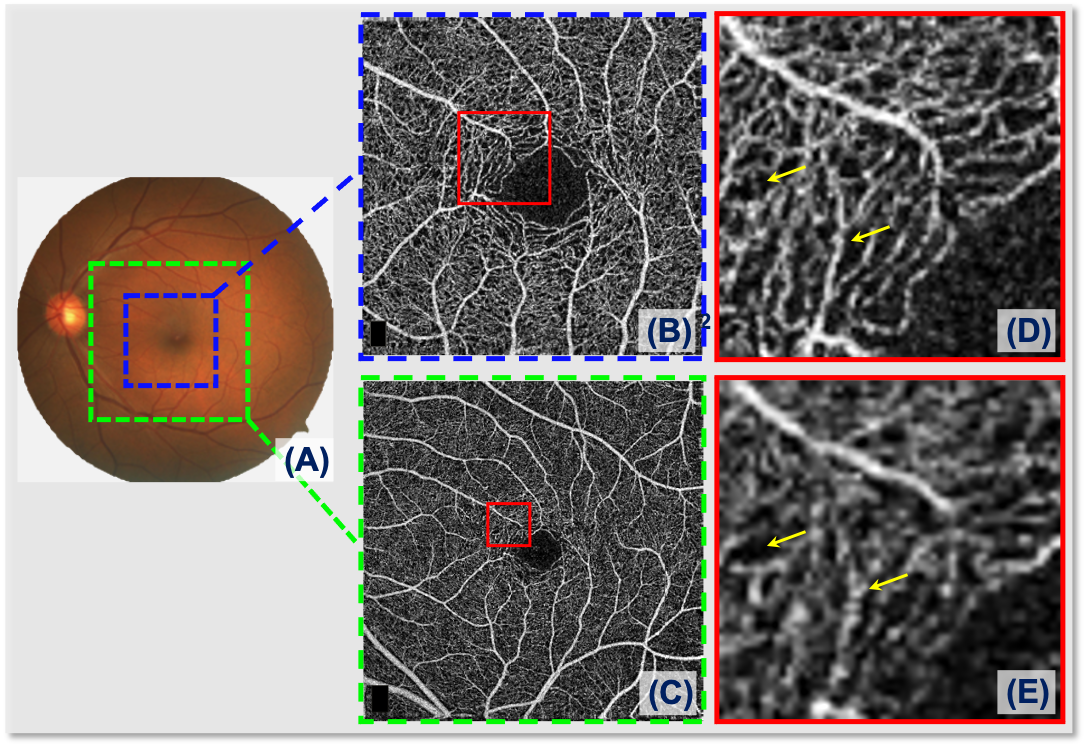}
}
\caption{Example of a (A) retinal color fundus image and its corresponding macula-centre (B) $3\times3~mm^2$ and (C) $6\times6~mm^2$ OCTA \textit{en face} images. (D) and (E) are two zoomed image patches of the same scan size ($1.0\times1.0~mm^2$) of (B) and (C), respectively. In the two image patches, there are local capillary differences at the locations indicated by the yellow arrows.}
\label{fig.1}
\end{figure}

\IEEEPARstart{O}{ptical} coherence tomography angiography (OCTA) is a non-invasive imaging technique that is widely used for retinal vascular imaging. Unlike conventional vascular imaging techniques such as fluorescein angiography (FA), OCTA successfully avoids the potential side effects and risks associated with dye injection~\cite{de2015review}. Compared to color fundus imagery, as shown in  Fig.~\ref{fig.1} (B), OCTA is able to capture the  microvasculatures surrounding in the fovea and parafovea regions~\cite{spaide2015image}. Therefore, it is often used to quantify important clinical indicators, such as vessel density, curvature and fractal dimension, to assist clinicians in the diagnosis and treatment of retina-related diseases~\cite{MouMICCAI,delia2017retinal,eladawi2017automatic}.
However, the diagnosis of some retinal diseases often requires the analysis of vascular structure from a larger field of view (FOV). For example, Diabetic Retinopathy (DR) mainly presents with macular peripheral vascular changes, but some important vessels are often not visible in OCTA imaging scans due to the limitation of imaging range~\cite{you2020detection}. Meanwhile, the resolution of OCTA images is inversely proportional to the field of view at the same sampling frequency, which limits clinicians' performance for analyzing larger vascular areas. In consequence, an OCTA image with high resolution is highly desired when dealing with larger FOV conditions. 

Current OCTA imaging devices, such as the RTVue XR Avanti SD-OCT system (Optovue, USA) and CIRRUS HD-5000 (Carl Zeiss, Germany) can produce OCTA \textit{en face} images with different field of views, e.g.,  $3\times3~mm^2$,  $6\times6~mm^2$, and $8\times8~mm^2$. The $3\times3~mm^2$ and $6\times6~mm^2$ scans are two commonly used scanning approaches in clinical practice. The former has a higher scanning density, which means it has a higher image resolution and can depict the retinal foveal avascular zone (FAZ) and surrounding capillaries more clearly, as shown in  Fig.~\ref{fig.1} (B) and (D). By contrast, as shown in Fig.~\ref{fig.1} (C) and (E), the $6\times6~mm^2$ scan has a larger FOV but with lower scanning density, which results in lower imaging visibility to present small  capillaries~\cite{ma2020rose}.  This is due to the fact that a high scanning/sampling frequency is hardly to be achieved in clinical practice by considering its longer inter-frame and total imaging time at a larger FOV. Artifacts caused by involuntary movements, blinking and tear film evaporation will arise during the long acquisition procedure and interfere with clinical diagnosis~\cite{wei2020high}. Thus, it has been becoming a challenging issue to ensure a considerable image resolution with a larger FOV. 

In recent years, several works have been established to achieve high imaging resolution within larger FOV images by improving the optical system directly~\cite{wieser2010multi}. Despite the significant efforts have been invested, the current state-of-the-art OCTA acquisition systems still need to make a trade-off between image resolution and FOV size~\cite{wei201975}. Hence, pre-processing techniques such as image enhancement is elaborated to compensate the resolution loss and obviously show more values.  For example, Zang \textit{et al.}~\cite{zang2016automated} proposed to minimize image artifacts in order to extend the total samplings and thus to obtain high resolution images with larger FOV.  Uji \textit{et al.}~\cite{uji2018multiple} employed a multiple \textit{en face} image averaging method to  enhance image quality of OCTA; Tan \textit{et al.}~\cite{tan2018enhancement} developed a vessel enhancement algorithm based on a modified Bayesian residual transform, which can improve the contrast and visibility of vascular features. However, such procedures may require more repeated acquisitions, which in turn increases the total sampling time. 


Image super-resolution reconstruction algorithms have been widely applied in natural and medical images~\cite{wang2015deep,sun2008image,yang2013fast,yang2010image,dong2015image,umehara2018application,shi2018super}, as they are not only able to perform HR reconstruction of images with the larger FOVs, but also reduce the necessity of long acquisition time. Currently, image super-resolution techniques can be divided into three types: interpolation-based~\cite{wang2015deep}, statistics-based~\cite{sun2008image,yang2013fast,yang2010image}, and learning-based methods~\cite{niu2020single,zhao2021large,wang2018esrgan}. Most of image super-resolution techniques are always limited by detailed, realistic textures or small upscaling factors, while learning-based methods are capable of recovering missing high-frequency information from large quantities of LR and HR pairs, and thus have achieved significant improvements in super-resolution reconstruction~\cite{dong2015image,umehara2018application,shi2018super,mahapatra2019image,feng2021multi}. 
For example, By introducing Super-Resolution Convolutional Neural Network (SRCNN)~\cite{dong2015image}, convolutional neural networks (CNNs) have led to state-of-the-art performance in super-resolution reconstruction of medical images. 
Umehara \textit{et al.}~\cite{umehara2018application} proposed to apply SRCNN on CT images and their approach outperforms many of the traditional linear interpolation methods.
For MRI images, Shi~\textit{et al.}~\cite{shi2018super} developed a residual learning-based super-resolution method to solve MRI super-resolution problems. 
Feng \textit{et al.}~\cite{feng2021multi} proposed a multi-stage integration network for
multi-contrast MRI super-resolution, which explicitly models the dependencies between multi-contrast images at different stages.
Zhou \textit{et al.}~\cite{zhou2019image} proposed a two-stage GAN to improve the reconstruction of ultrasound image structures and details by cascading a U-Net network at the front end of the generator and a modified GAN network at the back end. The network combined multi-scale global residual and local residual learning, which can effectively capture high-frequency details. These successes have motivated our investigation of OCTA super-resolution methods using CNNs. Hence, in this paper, we aim to enhance the image quality of $6\times6~mm^2$ OCTA via learning-based super-resolution method for alleviating the resolution loss caused by under-sampling. 


In general, learning-based super-resolution methods require pairs of LR-HR images. In practice, it is extremely difficult to obtain a realistic HR image in $6\times6~mm^2$ OCTA acquisition. 
Since $3\times3~mm^2$ OCTA images have higher resolution and are also fully contained within the $6\times6~mm^2$ images, we can therefore use realistic $3\times3~mm^2$ OCTA images as the reference HR images to guide the training of  realistic $6\times6~mm^2$ images reconstruction. This strategy requires the registration of $3\times3~mm^2$ to $6\times6~mm^2$. However, it is difficult to achieve perfect registration result, due to the fact that local capillary differences caused by eye motion and signal variations obviously exist between the two acquisitions, as shown in Fig.~\ref{fig.1}. Such discrepancy in the spatial mapping between $3\times3~mm^2$ and $6\times6~mm^2$ images has been observed in several OCTA repeatability and reproducibility studies~\cite{hong2020intra,lei2017repeatability,men2017repeatability,lee2019repeatability}. Ignoring this domain discrepancy during learning process could result in over-smoothed reconstruction with less detailed information. It is a key aspect to reduce the domain gap between $3\times3~mm^2$ and $6\times6~mm^2$ images for high-resolution vascular structures reconstruction.

Domain adaptation aims to use labeled source domains to learn models that perform well on unlabeled target domains. 
Recently, adversarial-based domain adaptation methods have been proposed for solving challenging dense prediction tasks~\cite{sankaranarayanan2017unsupervised,zhang2017curriculum,tsai2018learning}.
For example, Tsai  \textit{et al.}~\cite{tsai2018learning} proposed an adversarial-based domain adaptation method for semantic segmentation, which uses the spatial similarity of the source and target domain outputs to reduce the domain differences. By means of appropriate adaptation strategies, models trained on synthetic datasets have achieved performance comparable to that of models trained with realistic labeled datasets.~\cite{bousmalis2017unsupervised}.

Inspired by these adversarial-based domain adaptation methods, we propose a novel framework from a domain adaptation perspective for $6\times6~mm^2$ OCTA images super-resolution reconstruction, which can mitigate the difficulties in pairwise training of $3\times3~mm^2$ and $6\times6~mm^2$ images. In the proposed network, a multi-level super-resolution model is proposed as a generator for $6\times6~mm^2$ images reconstruction, and PatchGAN~\cite{li2016precomputed} is developed as a discriminator to align the reconstruction results with the features of HR images. In addition, we introduce a sparse edge-aware loss to optimize the vascular structure of the reconstruction results via dynamic alignment of edges between the reconstructed HR image and the reference HR image. Finally, we perform quality evaluation of the super-resolution results on two OCTA sets with vascular and FAZ pixel annotations. The experimental results show that our proposed method achieves state-of-the-art results in terms of visual quality and clinical evaluation on two OCTA sets.

\begin{figure}[t]
\centering{
\includegraphics[width=8.5cm]{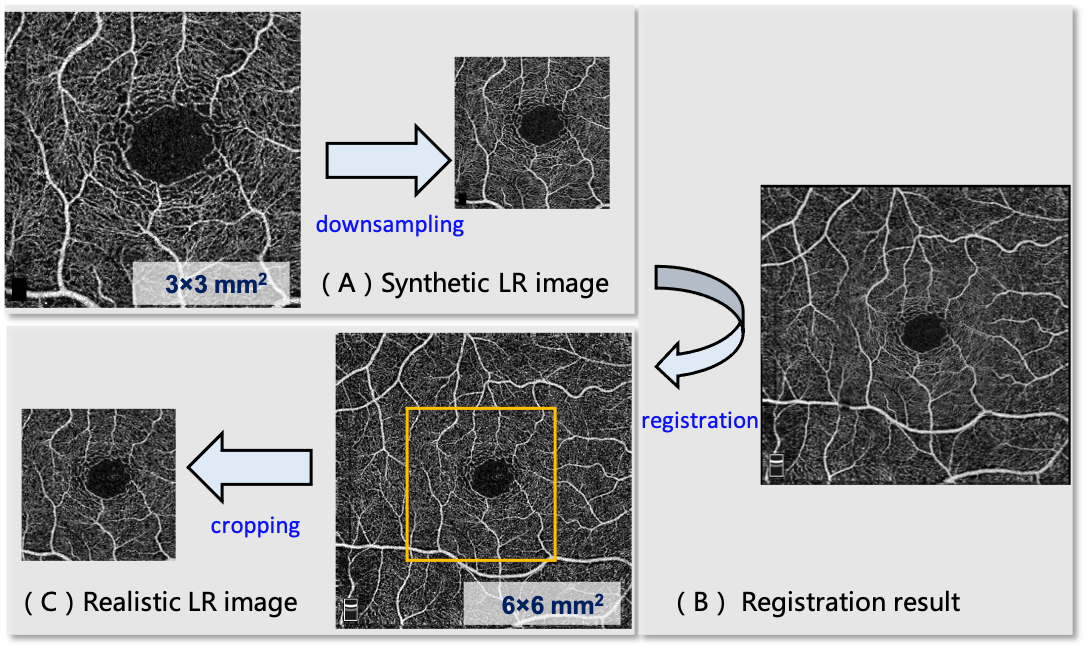}
}
\caption{Demonstration of  (B) registration result between $3\times3~mm^2$ and $6\times6~mm^2$ images and LR image generation by two different ways: (A) Synthetic LR image and (C) realistic LR image generations.}
\label{fig.3}
\end{figure}

Our contributions can be summarized as follows:

$\bullet$ An adversarial-based super-resolution model is designed in this work. It aligns the spatial features between the generated HR image and the reference HR image to address the domain gap between LR images and its reference HR images.

$\bullet$ We introduce a sparse edge-aware loss to dynamically optimize the reconstruction of local structures in LR images. This new loss provides the flexibility to learn similar edge features from the reference HR images.

$\bullet$ The proposed method has undergone rigorous quantitative and qualitative evaluation, and it has demonstrated the proposed method has promoted the OCTA image segmentation performance.
\vspace{1em}
\section{Dataset}
In this work, a new \textbf{SU}per-resolution \textbf{RE}construction dataset (\textbf{SURE}) of OCTA image is constructed for the proposed work. 
SURE includes two subsets: SURE-O and SURE-Z, and the images in these two sets were acquired by two commonly-used OCT imaging devices: \textbf{\underline{O}}ptovue Avanti RTVue XR (Optovue, Fremont, USA), \textbf{\underline{Z}}eiss Cirrus HD-OCT 5000 (Zeiss Meditec, Dublin, USA), respectively.

$\bullet$ \textbf{SURE-O}  contains 559 pairs of OCTA images from 320 subjects (including 150 with Alzheimer’s disease (AD) and 170 healthy controls) with FOV of $3\times3~mm^2$ and $6\times6~mm^2$ separately. Both FOVs have the same image resolution of $304\times304$. We randomly assigned the 559 pairs of images to the training, testing, and validation sets in a 4:1:1 ratio.
 
$\bullet$ \textbf{SURE-Z}  contains 261 pairs of OCTA images from 172 subjects (including 159 with diabetic retinopathy (DR) and 13 healthy controls) with FOVs of $3\times3~mm^2$ and $6\times6~mm^2$ separately. Both FOVs have the same image resolution of $512\times512$. Similar to the SURE-O dataset, we divided the SURE-Z into the training, testing, and validation sets in the ratio of 4:1:1.

Super-resolution reconstruction task usually requires aligned image pairs in the training phase. To this end, we separately have a pair of \textit{synthetic} LR-to-HR images and a pair of \textit{realistic} LR-to-HR images in this work. We define the original $3\times3~mm^2$ OCTA images from both sets as HR images. The corresponding low resolution images were obtained by two different strategies. First, the original $3\times3~mm^2$ OCTA images were downsampled to obtain \textbf{Synthetic LR} image (${D}_{LR}$). The original image was degraded from  $304\times304$ to $152\times152$, and we use bicubic interpolation in this work. We then cropped the corresponding area $3\times3~mm^2$ scanned area from $6\times6~mm^2$ image, which is defined as \textbf{Realistic LR} image ${I}_{LR}$ and the Generalised Dual Bootstrap-ICP (GDB-ICP)~\cite{tsai2009edge} registration method is employed.  Fig.~\ref{fig.3} illustrates the strategies of LR image generations.

In addition, we use retinal vessel and FAZ segmentation to better confirm the impact of proposed reconstruction model on subsequent analysis tasks. Three well-trained imaging experts manually labeled the retinal vessel and FAZ, and then two senior ophthalmologists reviewed and refined the manual annotations.  Their consensus were finally defined as ground truth in this study. The inter-annotator agreement is higher than 0.90 in terms of pixel-level. 37 and 36 OCTA images (all with $3\times3~mm^2$ scan) from the testing set of  SURE-O and SURE-Z were randomly selected for annotation, respectively. All the data described in this section have appropriate approvals from the ethics committees of Ningbo Institute of Industrial Technology, Chinese Academy of Sciences, and written informed consent was obtained from each participant in accordance with the Declaration of Helsinki.

\section{Methodology}


As aforementioned, due to the noise and contrast variations in the realistic LR-to-HR pairs, it is challenging to obtain satisfactory reconstruction performance by solely relying on the realistic pairs to  supervise a pixel-by-pixel learning process. Therefore, the key of our network lies on how synthetic data pairs can guide the super-resolution reconstruction of realistic data pairs. To this end, we propose a semi-supervised framework which exploits the concept of domain adaptation reconstruction. The proposed Sparse-based domain Adaptation Super-Resolution (SASR) method consists of three components, i.e., a multi-level super-resolution network (MLSR), a discriminator network, and a sparse edge-aware loss function.  It is worth noting that we define the adversarial network including the MLSR and patch discrimination modules as a domain adaptation super-resolution network (DASR).
Fig.~\ref{fig.3} shows the overall network architecture. 

\begin{figure*}[!t]
\centering{
\includegraphics[width=18cm]{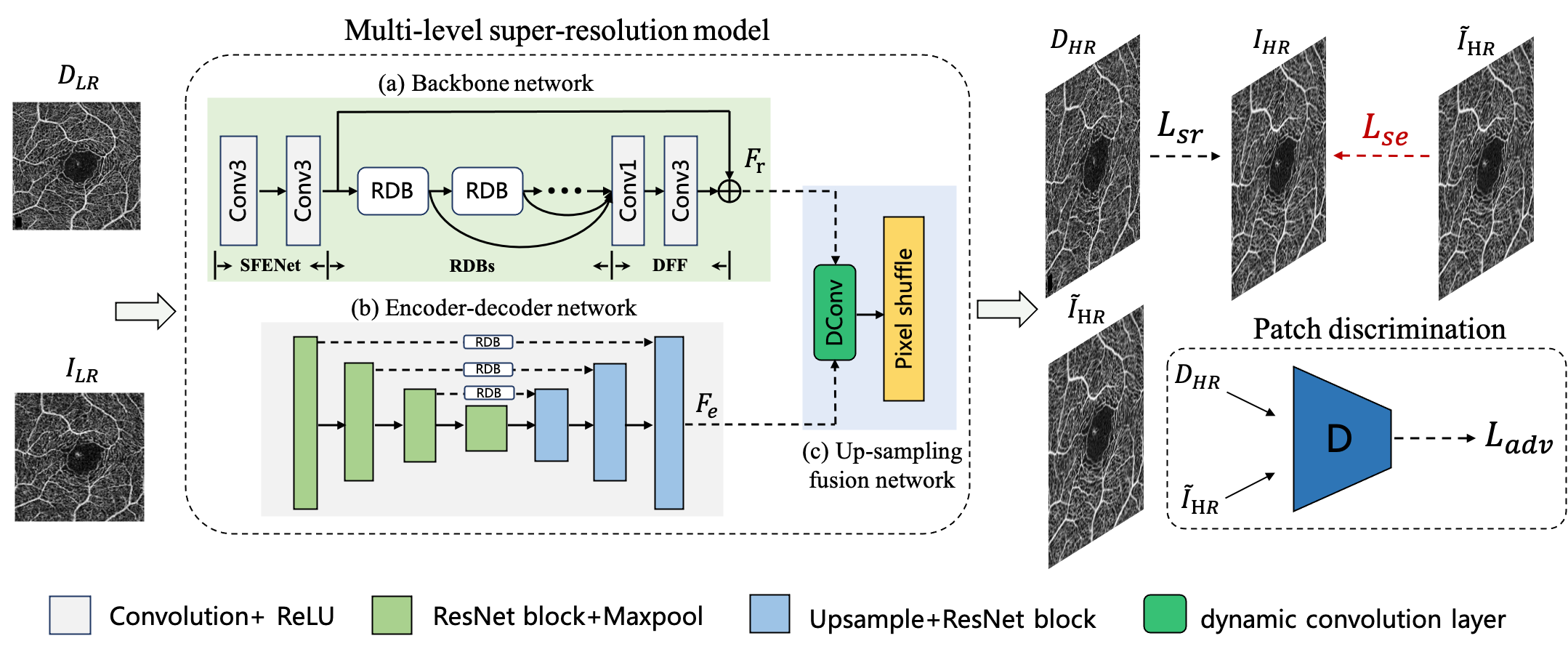}
}
\caption{Architecture of the proposed SASR network for OCTA image reconstruction. The proposed SASR network mainly consists of multi-level super-resolution model (MLSR), patch discrimination and sparse edge-aware loss ($\mathcal{L}_{\mathrm{se}}$). The MLSR is used as the generation model in the SASR, including (a) backbone network, (b) encoder-decoder network,  and (c) up-sampling fusion network.}
\label{fig.3}
\end{figure*}

\subsection{Multi-level super-resolution network}
Unlike image classification which is adaptive in the global feature space, adaptive learning in low-dimensional space is optimal for dense estimation tasks. Our intuition is that the  reconstructed image of $6 \times 6~mm^2$ OCTA has strong spatial and local similarities to the $3 \times 3~mm^2$ HR OCTA image. Therefore, we exploit this property to adapt the low-dimensional output of super-resolution reconstruction by means of generative-adversarial learning, in which a MLSR network is proposed as a super-resolution generative network and a patch discriminator as a discriminative network.

The MLSR network is introduced for the supervised super-resolution reconstruction of ${D}_{LR}$, while ${I}_{LR}$ and ${D}_{LR}$ share the network parameters to generate high-resolution reconstruction results of the realistic LR images. As shown in Fig.~\ref{fig.3}, the MLSR consists of three main components: a backbone network, an encode-decode network, and an up-sampling fusion network. The backbone network adopts the RDN~\cite{zhang2018residual} structure, which consists of four main parts: the shallow feature extraction net (SFENet), residual dense blocks (RDBs), dense feature fusion (DFF), and the up-sampling net (UPNet). In our method, we remove the UPNet structure from the RDN network and preserve the framework of SFENet, RDBs and DFF. 
As shown in Fig.~\ref{fig.3} (a), SFENet consists of two $3\times3$ convolutional layers, and RDBs are composed of six dense residual modules and one $1\times1$ convolutional layer, forming a continuous memory mechanism.
Contiguous memory mechanism means dense connection, which is realized by passing the state of preceding RDB to each layer of current RDB. This mechanism ensures continuous storage and memory of low-level and high-level information.
We denote the output of this module as ${F}_{r}$.

The backbone network has extracted the dense multi-scale features of the image. However, the multi-resolution features are not fully utilized by the network. Therefore, in the parallel sub-network, we introduce an encoder-decoder network to extract the multi-resolution features of the image and optimize the overall structure. This sub-network consists of three encoder-decoder layers with the skip-connection block. As can be seen in Fig.~\ref{fig.3} (b), each green box represents the encoder layer, which includes two residual modules and a maxpooling layer. The residual module contains three $3\times3$ convolutional layers and three batch normalization layers. The blue boxes represent the decoder layers, and each of them includes an up-sampling layer and a residual module, where the up-sampling layer employs the nearest neighboring interpolation method. 
The white boxes represent the RDB modules, which is formed by combining a residual block and a dense connection block to enhances the transmission of information and gradients.
We add the RDB module to the skip connection with the aim of improving the ability to extract and transfer spatial information to the input feature map. The output of encoder-decoder network is denoted by ${F}_{e}$. 

\begin{figure}[t]
\centering{
\includegraphics[width=8cm]{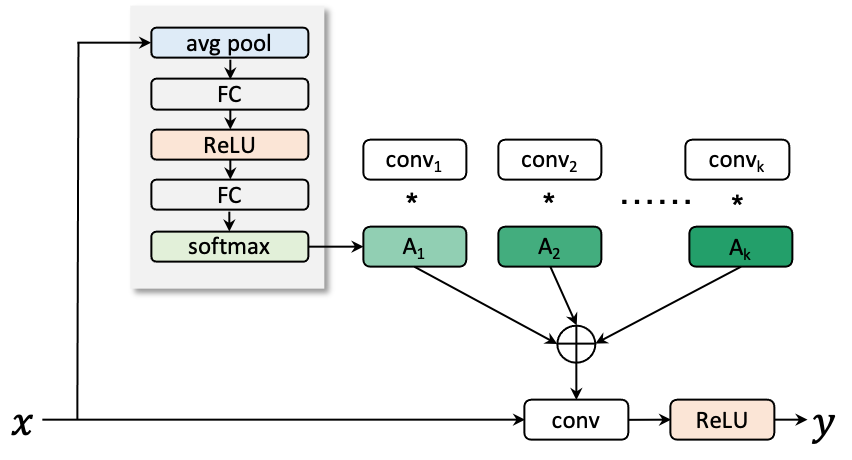}
}
\caption{Architecture of the dynamic convolution.}
\label{fig.4}
\end{figure}

Afterwards, we concatenate ${F}_{r}$ and ${F}_{e}$ into a new feature map $x$ as the input to the up-sampling fusion network. 
This network mainly consists of a dynamic convolution layer and an up-sampling layer. The dynamic convolution~\cite{chen2020dynamic} can guide the layer parameters adaptively updated according to the feature maps of the different input channels, which can synthesize and optimize the reconstruction results of the output at different stages. 
As shown in Fig.~\ref{fig.4}, the weight matrix $W \in \mathbb{R}^{ K \times N_\text{in} \times N_\text{out} \times d \times d}$ for each dynamic convolutional layer, where $K$, $N_\text{in}$, $N_\text{out}$ and $d$ are the number of kernels, input channel, output channel and kernel size, can be defined as:
\begin{equation}\small
\begin{aligned} 
{{W}}({x})=\sum_{k=1}^{K} {A}_{k}({x}) {{W}}_{k}, 
\end{aligned}
\end{equation}
\begin{equation}\small
\begin{aligned} 
 {A}(x) = \text{Softmax}(MLP(GAP(x))), \\
\end{aligned}
\end{equation}
\begin{equation}\small
\begin{aligned} 
\sum_{k=1}^{K} A_{k}(\boldsymbol{x})=1, \quad 0 \leq A_{k}(\boldsymbol{x}) \leq 1, 
\end{aligned}
\end{equation}
where ${{W}}_{k}$ and $A_{k}$ represent the weight matrix and attention map of the $k^{th}$ kernel, and we set $K=4$ in the experiments.  Note that $A_{k}$ is obtained by using a Squeeze and Excitation (SE) module~\cite{hu2018squeeze} to obtain discriminative representation. More specifically, the SE module is composed of a global average pooling (GAP) layer, and a Multi-Layer Perceptron (MLP) with a ReLU-activated hidden layer, followed by the Softmax layer. The output of ${W}({x})$ is denoted as $y$, which is fed into the PixelShuffle~\cite{huang2009multi} layer to obtain the high-resolution reconstruction results of the same size as  ${I}_{HR}$. 
The PixelShuffle layer first feeds the low-resolution feature map into a convolutional layer for channel expansion, and then performs multi-channel reorganisation to generate a high-resolution map through period filtering. Compared to other upsampling methods, the PixelShuffle is able to increase the frequency of information extraction to enrich the image detail texture.
In the training stage, we combine the Mean Squared Error (MSE) and SSIM as the super-resolution loss $\mathcal{L}_{sr}$ to optimize the output in this network. The loss can be defined as:
\begin{equation}
\mathcal{L}_{sr}=\lambda_{\mathrm{MSE}} \cdot \mathcal{L}_{\mathrm{MSE}}+\lambda_{\mathrm{SSIM}} \cdot \mathcal{L}_{\mathrm{SSIM}}
\end{equation}
where $\lambda_{\mathrm{MSE}}$ and $\lambda_{\mathrm{SSIM}}$ are set to 1 and 0.5 empirically.
\begin{figure}[t]
\centering{
\includegraphics[width=8cm]{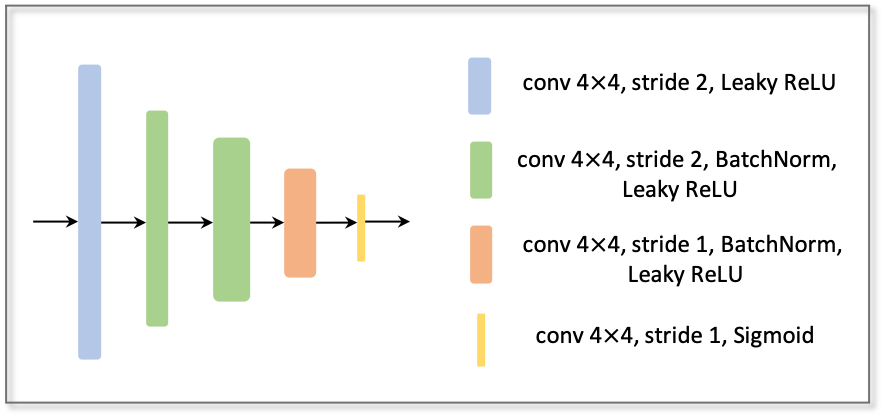}
}
\caption{The network structure of the patch discrimination.}
\label{fig.5}
\end{figure}
\subsection{Patch discrimination} 
To guide the realistic LR-to-HR image reconstruction, we use a discriminator to assist the generative network to achieve spatial feature alignment between synthetic and realistic data. 
In our paper, we use PatchGAN to classify true and false samples for synthetic and real LR image reconstruction results. Compared with traditional discriminators, PatchGAN makes judgments based on the patch level rather than the whole image level.
As shown in Fig.~\ref{fig.5}, the structure of PatchGAN is composed of $4\times4$ five convolutional layers with a stride of 2 in the first three layers and a stride of 1 in the last two layers. Except for the last convolutional layer, the first four convolutional layers are followed by Leaky ReLU layers with a slope of 0.2 and batch normalization (BN) layers. The Sigmoid layer is used to obtain the probability score of the output feature map in the last convolutional layer. In the above setting, the receptive field size of PatchGAN is $70\times70$, which means that PatchGAN has a faster inference speed than traditional discriminators, but still can guide the generator to generate realistic results. Consequently, each pixel in the final output feature map represents the probability that the corresponding $70\times70$ patch of the input image is from a real sample.
To this end, the loss function in the adversarial process of discriminator and generator is defined as:
\begin{equation}
\mathcal{L}_{\mathrm{adv}}=\mathbb{E}\left[\log \left(1-D\left(G({I}_{LR})\right)\right)\right]+\mathbb{E}[\log D({G({D}_{LR})})]
\end{equation}
where $D$ and $G$ are discriminator and MLSR, respectively.
\begin{figure}[t]
\centering{
\includegraphics[width=8.5cm]{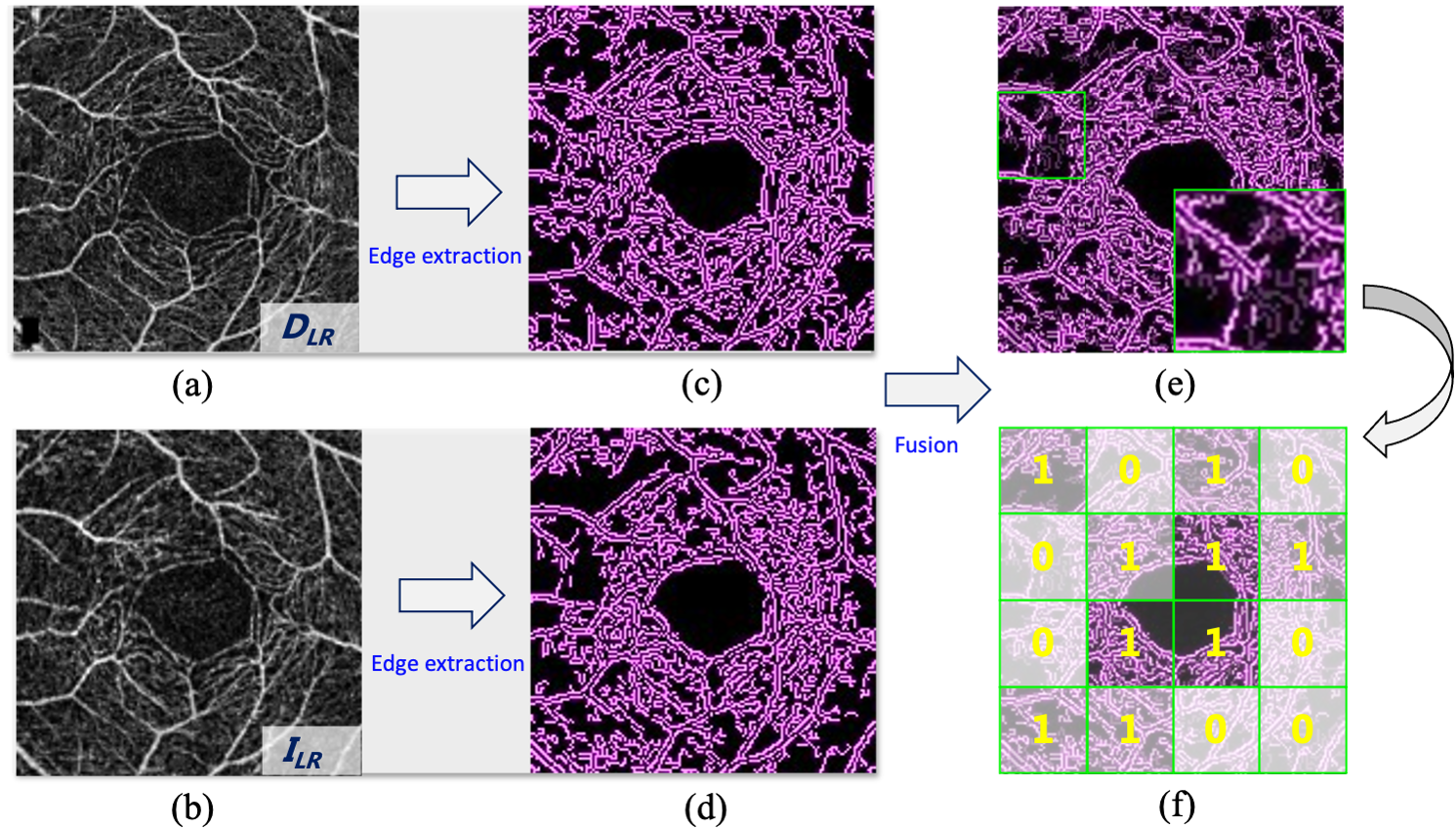}
}
\caption{Demonstration of sparse result between $D_{LR}$  and $I_{LR}$. (a) Synthetic LR image ($D_{LR}$). (b) Realistic LR image ($I_{LR}$). (c)-(d) Edge structures of (a) and (b) extracted by canny operator. (e) Fusion result of (c) and (d). (f) Sparse results determined by the similarity of each pair of patches in (e).}
\label{fig.6}
\end{figure}
\begin{figure*}[t]
\centering{
\includegraphics[width=18cm]{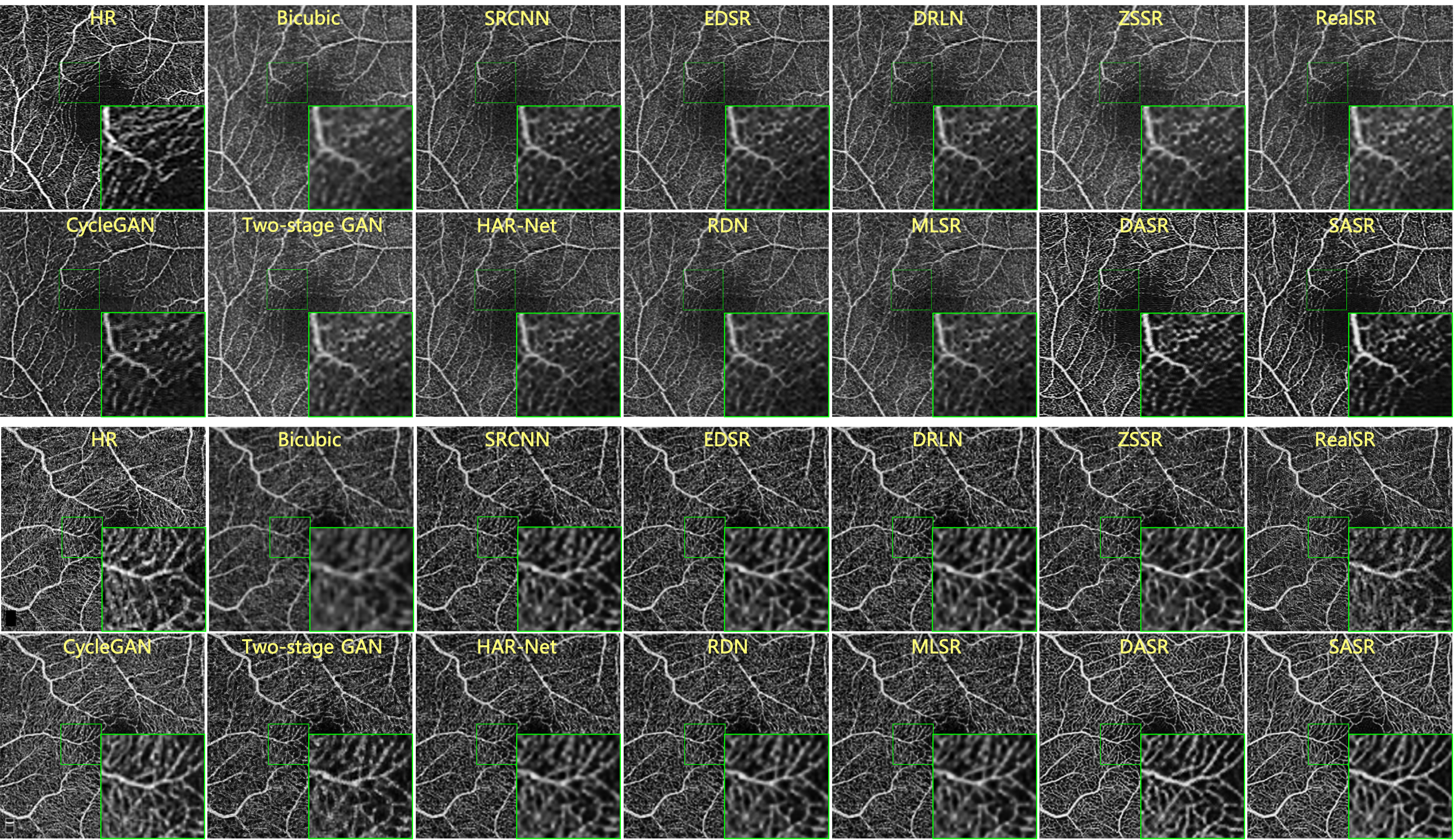}
}

\caption{Visual comparison of super-resolution images reconstructed by different methods over two randomly selected images from SURE-O and SURE-Z sets, respectively. Green rectangles indicate two representative patches for illustration.}
\label{fig.7}
\end{figure*}
\subsection{Sparse edge-aware loss}  
Unlike the general domain adaptation approach, the inputs to our proposed framework are paired synthetic and realistic LR images. Even though ${I}_{LR}$ and ${I}_{HR}$ have some discrepancy in spatial mapping, their vascular structures are very similar. In order to further improve the local vascular reconstruction of realistic LR images, we propose to use edge similarity loss for structure optimization. However, since the background noise affects the extraction of partial vessel structures, optimizing the overall vessel edges would introduce mislabeling and thus generate over-smoothed structures. To solve this issue, a sparse edge-aware loss is designed to adaptively constrain the reconstruction results. 
Firstly, as shown in Fig.~\ref{fig.6}, we use the canny operator~\cite{bao2005canny} to extract the edge structures of the ${I}_{LR}$ and ${D}_{LR}$.  Then MSE is employed to compute the distances between the two edge images as follows:

\begin{equation}
d(t) = \left \| C({D}_{t})-C({I}_{t})\right \|_{2}
\end{equation}
where ${D}_{t}$ and ${I}_{t}$ denote the ${t}_{th}$ patches with the size of $n \times n$  from ${D}_{LR}$ and ${I}_{LR}$ with $\mathrm{t}=1,2, \cdots, \mathrm{T}$, respectively. $d(t)$ denotes  the distance of each pair of patches. Then we adopt a hard shrinkage operation to promote the sparsity:
\begin{equation}
\label{eq8}
\widehat{w}_{t}=\frac{\max \left(d_{t}-\lambda, 0\right) }{\left|d_{t}-\lambda\right|+\epsilon}
\end{equation}
where $\max (\cdot, 0)$ is also known as ReLU activation, and $\epsilon$ is a very small positive scalar.  $\lambda$ is used  to screen vascular regions of ${I}_{LR}$ with structures similar to ${D}_{LR}$. Finally, we compute the edge distances between realistic LR image reconstruction result $\tilde{I}_{HR}$ and label ${I}_{HR}$:
\begin{equation}
\widehat{d}(t) = \left \| C({H}_{t})-C(\tilde{H}_{t})\right \|_{2}
\end{equation}
where ${H}_{t}$ and $\tilde{H}_{t}$ denote the ${t}_{th}$ patches with the size of $n \times n$  from ${I}_{HR}$ and $\tilde{I}_{HR}$. Therefore, the sparse edge-aware loss is defined as:
\begin{equation}
\mathcal{L}_{\mathrm{se}}= \sum_{t=1}^{T}\widehat{d}(t)\widehat{w}_{t}
\end{equation}
To this end, the total loss function is denoted as:
\begin{equation}
\mathcal{L}_{total}=\lambda_{\mathrm{MSE}} \cdot \mathcal{L}_{\mathrm{MSE}}+\lambda_{\mathrm{SSIM}} \cdot \mathcal{L}_{\mathrm{SSIM}}+\lambda_{\mathrm{adv}} \mathcal{L}_{\mathrm{adv}}+\lambda_{\mathrm{se}}\mathcal{L}_{\mathrm{se}}
\end{equation}
where $\lambda_{\mathrm{adv}}$ and  $\lambda_{\mathrm{se}}$ are set as 1 and 0.1 in our paper.

\section{Experiments}

\subsection{Implementation details}
The proposed method was implemented by the publicly available Pytorch Library in the Nvidia GeForce TITAN Xp GPU. In the training phase, we employed an Adam optimizer~\cite{kingma2014adam} to optimize the deep model. We used a gradually decreasing learning rate, starting from 0.0001, and a momentum of 0.9. In each iteration, we took a random $96 \times 96$ patch of the image for training, and the batch size was set to 8 during the training. We trained the network for 300 epochs and the network reached convergence at around the 150 epoch. The training lasted approximately 20 and 8 hours on the \textbf{SURE-O} and \textbf{SURE-Z} datasets, respectively.
In addition, online data enhancement with a random rotation from $-10^{\circ}$ to $10^{\circ}$ was employed to enlarge the training set.  In the sparse edge-aware loss, the parameter $\lambda$ in Equation~(\ref{eq8}) was set to 0.05 by counting the average distance between the edges of ${D}_{LR}$ and ${I}_{LR}$. $n$ and $T$ are set to 16 and 36, respectively.

\begin{table*}[t]
	\centering
	\caption{\textbf{FAZ} segmentation performance between the reconstructed and the reference HR images on the \textbf{SURE-O} dataset.}
	\setlength{\tabcolsep}{2.5mm}{
		\begin{tabular}{lllllllll} 
			\hline
			\multicolumn{1}{l||}{} & \multicolumn{7}{c}{\textbf{SURE-O}}\\
			\hline
				\multicolumn{1}{l||}{Methods}& \multicolumn{1}{c|}{\textit{AUC}  $\uparrow$}&
			\multicolumn{1}{c|}{\textit{Sen} $\uparrow$}&\multicolumn{1}{c|}{\textit{G-mean} $\uparrow$} &\multicolumn{1}{c|}{\textit{Kappa} $\uparrow$}
			&\multicolumn{1}{c|}{\textit{FDR} $\downarrow$}  &  
			\multicolumn{1}{c|}{\textit{Dice} $\uparrow$}
			&\multicolumn{1}{c}{\textit{$d_\text{HD}$}  $\downarrow$}& \\
			\hline 
			\multicolumn{1}{l||}{HR} &  \multicolumn{1}{c|}{0.979±0.013}
			& \multicolumn{1}{c|}{0.930±0.038} & 
			\multicolumn{1}{c|}{0.964±0.020}  &\multicolumn{1}{c|}{0.944±0.022} &\multicolumn{1}{c|}{0.036±0.025}&   \multicolumn{1}{c|}{0.946±0.021} &\multicolumn{1}{c}{15.58±23.13}\\
			\hline
			
			\multicolumn{1}{l||}{Bicubic~\cite{keys1981cubic}} &  \multicolumn{1}{c|}{0.945±0.068}&
			\multicolumn{1}{c|}{0.849±0.164} 
			& \multicolumn{1}{c|}{0.914±0.106}  &\multicolumn{1}{c|}{0.854±0.207}&   \multicolumn{1}{c|}{0.126±0.226} &\multicolumn{1}{c|}{0.859±0.199}
			&\multicolumn{1}{c}{29.28±40.31}\\
			
			\multicolumn{1}{l||}{ScSR~\cite{yang2010image}} &  \multicolumn{1}{c|}{0.947±0.064}& 
			\multicolumn{1}{c|}{0.856±0.147} 
			& \multicolumn{1}{c|}{0.919±0.092}  &\multicolumn{1}{c|}{0.866±0.183}
			&\multicolumn{1}{c|}{0.110±0.203}&   \multicolumn{1}{c|}{0.870±0.177}&  \multicolumn{1}{c}{24.51±30.66}  \\

			\multicolumn{1}{l||}{SRCNN~\cite{dong2015image}} &  \multicolumn{1}{c|}{0.939±0.079}&
			\multicolumn{1}{c|}{0.834±0.170} 
			& \multicolumn{1}{c|}{0.905±0.111}  &\multicolumn{1}{c|}{0.854±0.205}
			&\multicolumn{1}{c|}{0.112±0.224}&   \multicolumn{1}{c|}{0.859±0.198}&  \multicolumn{1}{c}{20.89±29.33}  \\
			
			\multicolumn{1}{l||}{EDSR~\cite{lim2017enhanced}} &  \multicolumn{1}{c|}{0.944±0.072}& 
			\multicolumn{1}{c|}{0.849±0.160} & 
			\multicolumn{1}{c|}{0.915±0.101} &  \multicolumn{1}{c|}{0.866±0.185}& \multicolumn{1}{c|}{0.106±0.198} &\multicolumn{1}{c|}{0.870±0.179} &
			\multicolumn{1}{c}{29.26±31.14} \\
			
			\multicolumn{1}{l||}{DRLN~\cite{anwar2020densely}} &  \multicolumn{1}{c|}{0.943±0.075}&
			\multicolumn{1}{c|}{0.851±0.159} 
			& \multicolumn{1}{c|}{0.916±0.101} & \multicolumn{1}{c|}{0.864±0.193}& 
			\multicolumn{1}{c|}{0.111±0.210}  & \multicolumn{1}{c|}{0.869±0.186} &\multicolumn{1}{c}{23.25±30.63}\\
				\multicolumn{1}{l||}{ RDN~\cite{zhang2018residual} (baseline)} &  \multicolumn{1}{c|}{0.942±0.076}& 
			\multicolumn{1}{c|}{0.846±0.166} & \multicolumn{1}{c|}{0.913±0.107}& \multicolumn{1}{c|}{0.863±0.192}&  
			\multicolumn{1}{c|}{0.110±0.205}&  \multicolumn{1}{c|}{0.867±0.186} 
			&\multicolumn{1}{c}{29.12±30.77} \\
			\hline
			
				\multicolumn{1}{l||}{ZSSR~\cite{shocher2018zero}} & 
				\multicolumn{1}{c|}{0.949±0.060}&
				\multicolumn{1}{c|}{0.857±0.152} & 
				\multicolumn{1}{c|}{0.919±0.095} & 
				\multicolumn{1}{c|}{0.869±0.183}& 
			\multicolumn{1}{c|}{0.108±0.199}  & 
			\multicolumn{1}{c|}{0.873±0.177} &
			\multicolumn{1}{c}{25.20±31.48}\\
			
			\multicolumn{1}{l||}{RealSR~\cite{ji2020real}} &  \multicolumn{1}{c|}{0.958±0.043}& 
			\multicolumn{1}{c|}{0.879±0.119} 
			& \multicolumn{1}{c|}{0.933±0.073} & \multicolumn{1}{c|}{0.868±0.170}& 
			\multicolumn{1}{c|}{0.128±0.195}  & \multicolumn{1}{c|}{0.873±0.164} &\multicolumn{1}{c}{29.83±36.22}\\
			
				\multicolumn{1}{l||}{CycleGAN~\cite{zhu2017unpaired}} &  \multicolumn{1}{c|}{0.942±0.027}&
			\multicolumn{1}{c|}{0.852±0.085} 
			& \multicolumn{1}{c|}{0.920±0.051} & \multicolumn{1}{c|}{0.849±0.118}& 
			\multicolumn{1}{c|}{0.142±0.139}  & \multicolumn{1}{c|}{0.854±0.114} &\multicolumn{1}{c}{34.22±32.11}\\
			
				\multicolumn{1}{l||}{Two-stage GAN~\cite{zhou2019image}} &  \multicolumn{1}{c|}{0.947±0.066}&
			\multicolumn{1}{c|}{0.852±0.165} 
			& \multicolumn{1}{c|}{0.916±0.106} & \multicolumn{1}{c|}{0.866±0.193}& 
			\multicolumn{1}{c|}{0.106±0.206}  & \multicolumn{1}{c|}{0.870±0.186} &\multicolumn{1}{c}{24.80±31.30}\\
		    
			\multicolumn{1}{l||}{HAR-Net~\cite{gao2020reconstruction}} &  \multicolumn{1}{c|}{0.942±0.076} &
			\multicolumn{1}{c|}{0.845±0.165} & \multicolumn{1}{c|}{0.912±0.106} & \multicolumn{1}{c|}{0.861±0.195}& 
			\multicolumn{1}{c|}{0.110±0.211}  & \multicolumn{1}{c|}{0.866±0.189}
			&\multicolumn{1}{c}{23.47±31.42}\\
			\hline

			\multicolumn{1}{l||}{SASR (Our Method)} &  \multicolumn{1}{c|}{$\textbf{0.961±0.034}$}& 
			\multicolumn{1}{c|}{$\textbf{0.890±0.093}$} & \multicolumn{1}{c|}{$\textbf{0.940±0.056}$} &   \multicolumn{1}{c|}{$\textbf{0.886±0.147}$}&  \multicolumn{1}{c|}{$\textbf{0.104±0.177}$}& \multicolumn{1}{c|}{$\textbf{0.890±0.142}$} &
			\multicolumn{1}{c}{{$\textbf{17.58±29.58}$}} 
			\\
			\hline
			
	\end{tabular}}
	\label{table1}
\end{table*}

\begin{table*}[t]
	\centering
	\caption{\textbf{FAZ} segmentation performance between the reconstructed and the reference HR images on the \textbf{SURE-Z} dataset.}
	\setlength{\tabcolsep}{2.5mm}{
		\begin{tabular}{lllllllll} 
			\hline
			\multicolumn{1}{l||}{} & \multicolumn{7}{c}{\textbf{SURE-Z}}\\
			\hline
			\multicolumn{1}{l||}{Methods}& \multicolumn{1}{c|}{\textit{AUC}  $\uparrow$ }&
			\multicolumn{1}{c|}{\textit{Sen} $\uparrow$}&\multicolumn{1}{c|}{\textit{G-mean} $\uparrow$} &\multicolumn{1}{c|}{\textit{Kappa} $\uparrow$}
			&\multicolumn{1}{c|}{\textit{FDR} $\downarrow$}  &  
			\multicolumn{1}{c|}{\textit{Dice} $\uparrow$}
			&\multicolumn{1}{c}{\textit{$d_\text{HD}$}  $\downarrow$}& \\
			\hline 
			\multicolumn{1}{l||}{HR} &  \multicolumn{1}{c|}{0.978±0.057}& 
			\multicolumn{1}{c|}{0.947±0.120} &
			\multicolumn{1}{c|}{0.970±0.068} &\multicolumn{1}{c|}{0.946±0.084}&  \multicolumn{1}{c|}{0.045±0.041}&  \multicolumn{1}{c|}{0.947±0.081}&  \multicolumn{1}{c}{28.27±44.70} \\
			\hline
			
			\multicolumn{1}{l||}{Bicubic~\cite{keys1981cubic}} &  \multicolumn{1}{c|}{0.916±0.051}& 
				\multicolumn{1}{c|}{0.763±0.114} 
				&\multicolumn{1}{c|}{0.867±0.069} &\multicolumn{1}{c|}{0.752±0.144}
			&   \multicolumn{1}{c|}{0.232±0.179} &\multicolumn{1}{c|}{0.760±0.139}&\multicolumn{1}{c}{105.38±93.92}\\
			
			\multicolumn{1}{l||}{ScSR~\cite{yang2010image}} &  \multicolumn{1}{c|}{0.932±0.039}& 
			\multicolumn{1}{c|}{0.809±0.092}
			&\multicolumn{1}{c|}{0.895±0.054}
			&   \multicolumn{1}{c|}{0.810±0.109} &\multicolumn{1}{c|}{0.170±0.134}
			&\multicolumn{1}{c|}{0.816±0.105} &\multicolumn{1}{c}{69.01±87.04}\\
			
			\multicolumn{1}{l||}{SRCNN~\cite{dong2015image}} &  \multicolumn{1}{c|}{0.937±0.028}&  
				\multicolumn{1}{c|}{0.817±0.068} 
				&\multicolumn{1}{c|}{0.901±0.039}
			&   \multicolumn{1}{c|}{0.846±0.072} &\multicolumn{1}{c|}{0.109±0.083}
			&\multicolumn{1}{c|}{0.851±0.069} &\multicolumn{1}{c}{74.37±39.51}\\
			
			\multicolumn{1}{l||}{EDSR~\cite{lim2017enhanced}} &  \multicolumn{1}{c|}{0.940±0.033}&    
				\multicolumn{1}{c|}{0.829±0.075}
				&\multicolumn{1}{c|}{0.908±0.043} 
			&\multicolumn{1}{c|}{0.848±0.083}
			&   \multicolumn{1}{c|}{0.118±0.093} &\multicolumn{1}{c|}{0.854±0.079}
			&\multicolumn{1}{c}{63.59±82.17}\\
			
			\multicolumn{1}{l||}{DRLN~\cite{anwar2020densely}} &  \multicolumn{1}{c|}{0.938±0.034}&  
				\multicolumn{1}{c|}{0.823±0.078} &\multicolumn{1}{c|}{0.904±0.045}&   \multicolumn{1}{c|}{0.839±0.092} 
			&\multicolumn{1}{c|}{0.129±0.108}&\multicolumn{1}{c|}{0.845±0.089}& 
			\multicolumn{1}{c}{86.30±86.87}\\
				\multicolumn{1}{l||}{ RDN~\cite{zhang2018residual} (baseline)} &  \multicolumn{1}{c|}{0.941±0.031}   &
				\multicolumn{1}{c|}{0.830±0.073} 
				&\multicolumn{1}{c|}{0.908±0.042}&  \multicolumn{1}{c|}{0.840±0.090} &\multicolumn{1}{c|}{0.134±0.110}   &\multicolumn{1}{c|}{0.846±0.087}&
			\multicolumn{1}{c}{69.92±87.87} \\
			\hline

				\multicolumn{1}{l||}{ZSSR~\cite{shocher2018zero}} &  \multicolumn{1}{c|}{0.927±0.042}&   
					\multicolumn{1}{c|}{0.797±0.093} &
				\multicolumn{1}{c|}{0.888±0.055} & 
				\multicolumn{1}{c|}{0.792±0.119}& 
			\multicolumn{1}{c|}{0.192±0.151}  & 
			\multicolumn{1}{c|}{0.799±0.114} &
			\multicolumn{1}{c}{82.25±88.34}\\

			\multicolumn{1}{l||}{RealSR~\cite{ji2020real}} &  \multicolumn{1}{c|}{0.932±0.043}& 
				\multicolumn{1}{c|}{0.813±0.097} &\multicolumn{1}{c|}{0.896±0.057}&   \multicolumn{1}{c|}{0.789±0.121} 
			&\multicolumn{1}{c|}{0.214±0.153}&\multicolumn{1}{c|}{0.794±0.117}& 
			\multicolumn{1}{c}{86.30±89.07}\\
			
				\multicolumn{1}{l||}{CycleGAN~\cite{zhu2017unpaired}} &  \multicolumn{1}{c|}{0.942±0.040}&  
				\multicolumn{1}{c|}{0.836±0.083} &\multicolumn{1}{c|}{0.912±0.048}&   \multicolumn{1}{c|}{0.878±0.073} 
			&\multicolumn{1}{c|}{$\textbf{0.065±0.059}$}&\multicolumn{1}{c|}{0.882±0.071}& 
			\multicolumn{1}{c}{49.82±40.89}\\
			
				\multicolumn{1}{l||}{Two-stage GAN~\cite{zhou2019image}} &  \multicolumn{1}{c|}{0.928±0.041}&  
				\multicolumn{1}{c|}{0.799±0.093} &\multicolumn{1}{c|}{0.889±0.055}&   \multicolumn{1}{c|}{0.791±0.116} 
			&\multicolumn{1}{c|}{0.194±0.147}&\multicolumn{1}{c|}{0.799±0.112}& 
			\multicolumn{1}{c}{63.13±85.29}\\
			
			\multicolumn{1}{l||}{HAR-Net~\cite{gao2020reconstruction}} &  \multicolumn{1}{c|}{0.940±0.035} &   
				\multicolumn{1}{c|}{0.829±0.810} 
				&\multicolumn{1}{c|}{0.907±0.047}  & \multicolumn{1}{c|}{0.840±0.095} & \multicolumn{1}{c|}{0.134±0.111}
			&\multicolumn{1}{c|}{0.845±0.091} & \multicolumn{1}{c}{85.55±85.32}\\	
					
			\hline
			
			\multicolumn{1}{l||}{SASR (Our Method)} &  \multicolumn{1}{c|}{$\textbf{0.958±0.033}$} &
				\multicolumn{1}{c|}{$\textbf{0.889±0.070}$} &\multicolumn{1}{c|}{$\textbf{0.940±0.039}$}&   \multicolumn{1}{c|}{$\textbf{0.897±0.063}$} &\multicolumn{1}{c|}{0.086±0.061}  
			&\multicolumn{1}{c|}{$\textbf{0.900±0.060}$} &\multicolumn{1}{c}{$\textbf{41.01±42.88}$}\\
			\hline
			
	\end{tabular}}
	\label{table2}
\end{table*}

\begin{figure*}[t]
\centering{
\includegraphics[width=18cm]{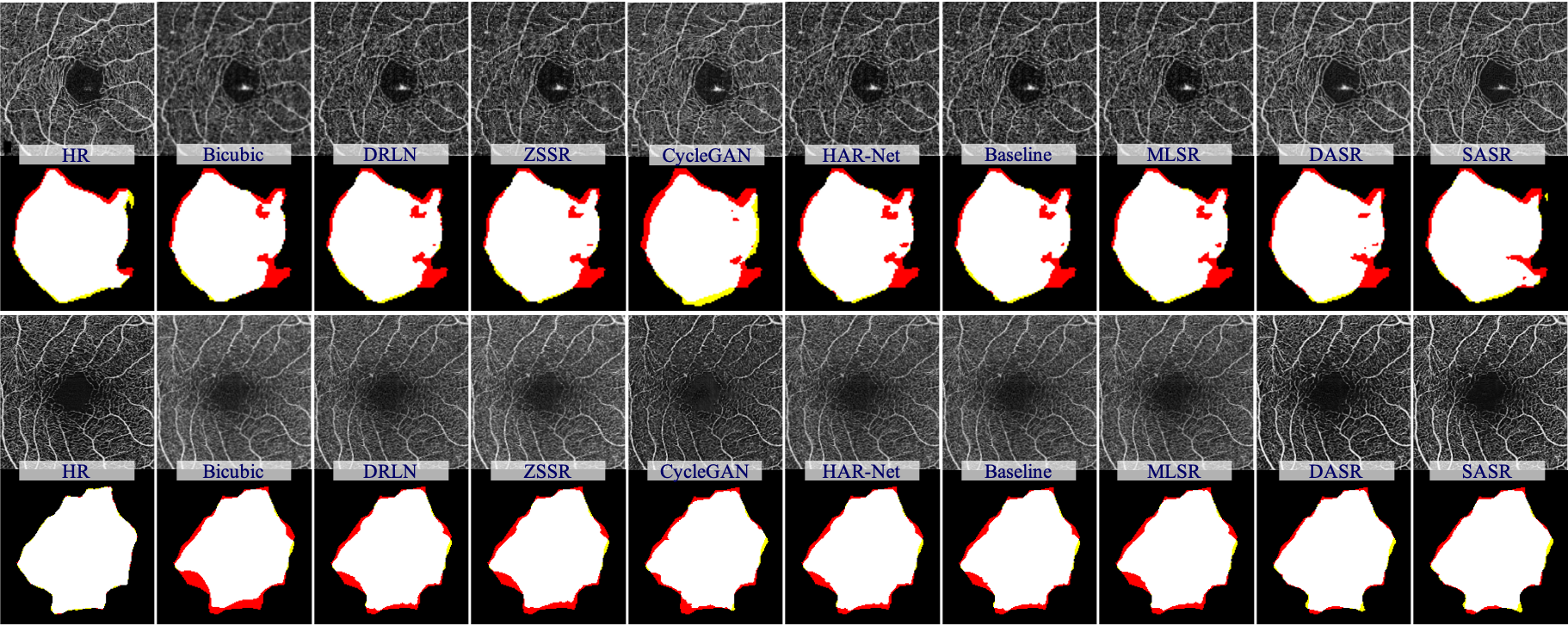}
}
\caption{FAZ segmentation performance on reconstructed OCTA image by different methods over two sample OCTA images selected from SURE-O and SURE-Z sets. Red represents under-segmentation, and blue indicates over-segmentation.}
\label{fig.8}
\end{figure*}

\begin{table*}[t]
	\centering
	\caption{\textbf{Vessel} segmentation performance between the reconstructed and the reference HR images on the \textbf{SURE-O} dataset. }
	\setlength{\tabcolsep}{2.5mm}{
		\begin{tabular}{lllllllll} 
			\hline
			\multicolumn{1}{l||}{} & \multicolumn{6}{c}{\textbf{SURE-O}}\\
			\hline
			\multicolumn{1}{l||}{Methods}& \multicolumn{1}{c|}{\textit{AUC}  $\uparrow$ }&
			\multicolumn{1}{c|}{\textit{Sen} $\uparrow$}&\multicolumn{1}{c|}{\textit{G-mean} $\uparrow$} &\multicolumn{1}{c|}{\textit{Kappa} $\uparrow$}
			&\multicolumn{1}{c|}{\textit{FDR} $\downarrow$}  &  
			\multicolumn{1}{c}{\textit{Dice} $\uparrow$} \\
			\hline 
			\multicolumn{1}{l||}{HR} &  \multicolumn{1}{c|}{0.891±0.027}&   
			\multicolumn{1}{c|}{0.718±0.067} &\multicolumn{1}{c|}{0.832±0.037}&  \multicolumn{1}{c|}{0.726±0.047}&  \multicolumn{1}{c|}{0.132±0.048}&  \multicolumn{1}{c}{0.783±0.041} \\
			\hline
			
			\multicolumn{1}{l||}{Bicubic~\cite{keys1981cubic}} &  \multicolumn{1}{c|}{0.788±0.029}&   \multicolumn{1}{c|}{0.526±0.054} &\multicolumn{1}{c|}{0.710±0.033}
			&   \multicolumn{1}{c|}{0.553±0.035} &\multicolumn{1}{c|}{$\textbf{0.196±0.075}$}&\multicolumn{1}{c}{0.631±0.036}\\
			
			\multicolumn{1}{l||}{ScSR~\cite{yang2010image}} &  \multicolumn{1}{c|}{0.791±0.033}&  
			\multicolumn{1}{c|}{0.569±0.048} &\multicolumn{1}{c|}{0.735±0.030}
			&   \multicolumn{1}{c|}{0.568±0.044} &\multicolumn{1}{c|}{0.229±0.087}&\multicolumn{1}{c}{0.651±0.042}\\
			
			\multicolumn{1}{l||}{SRCNN~\cite{dong2015image}} &  \multicolumn{1}{c|}{0.793±0.033}&  
	\multicolumn{1}{c|}{0.581±.0.044} &\multicolumn{1}{c|}{0.741±0.027}
			&\multicolumn{1}{c|}{0.571±0.046} &\multicolumn{1}{c|}{0.237±0.090}
			&\multicolumn{1}{c}{0.656±0.042}\\

			\multicolumn{1}{l||}{EDSR~\cite{lim2017enhanced}} &  \multicolumn{1}{c|}{0.792±0.032}&    
			\multicolumn{1}{c|}{0.575±0.048} &\multicolumn{1}{c|}{0.738±0.030}
			&   \multicolumn{1}{c|}{0.569±0.048} &\multicolumn{1}{c|}{0.234±0.094}&\multicolumn{1}{c}{0.653±0.044} \\

			\multicolumn{1}{l||}{DRLN~\cite{anwar2020densely}} &  \multicolumn{1}{c|}{0.794±0.033}&    
			\multicolumn{1}{c|}{0.579±0.046} &\multicolumn{1}{c|}{0.740±0.029}&   \multicolumn{1}{c|}{0.572±0.048} &\multicolumn{1}{c|}{0.233±0.091}
			&\multicolumn{1}{c}{0.655±0.044}\\
				\multicolumn{1}{l||}{ RDN~\cite{zhang2018residual} (baseline)} &  \multicolumn{1}{c|}{0.793±0.032}&  
			\multicolumn{1}{c|}{0.579±0.047} &\multicolumn{1}{c|}{0.740±0.029}
			&   \multicolumn{1}{c|}{0.571±0.048} &\multicolumn{1}{c|}{0.235±0.093}&\multicolumn{1}{c}{0.655±0.044} \\
			\hline
			
				\multicolumn{1}{l||}{ZSSR~\cite{shocher2018zero}} &  
				\multicolumn{1}{c|}{0.790±0.033}&   
				\multicolumn{1}{c|}{0.574±0.046} & 
				\multicolumn{1}{c|}{0.737±0.029}& 
			\multicolumn{1}{c|}{0.569±0.047}  & 
			\multicolumn{1}{c|}{0.233±0.093} &
			\multicolumn{1}{c}{0.652±0.044}\\
		
			\multicolumn{1}{l||}{RealSR~\cite{ji2020real}} &  
			\multicolumn{1}{c|}{0.784±0.030}&    
			\multicolumn{1}{c|}{0.556±0.047}&   
				\multicolumn{1}{c|}{0.725±0.029} 
			&\multicolumn{1}{c|}{0.554±0.040}
			&\multicolumn{1}{c|}{0.239±0.084}& 
			\multicolumn{1}{c}{0.639±0.041}\\
	
			\multicolumn{1}{l||}{CycleGAN~\cite{zhu2017unpaired}} &  \multicolumn{1}{c|}{0.791±0.036}&    
			\multicolumn{1}{c|}{0.589±0.048} &\multicolumn{1}{c|}{0.745±0.030}&   \multicolumn{1}{c|}{0.575±0.050} &\multicolumn{1}{c|}{0.241±0.093}
			&\multicolumn{1}{c}{0.659±0.047}\\
			
				\multicolumn{1}{l||}{Two-stage GAN~\cite{zhou2019image}} &  \multicolumn{1}{c|}{0.789±0.033}&    
			\multicolumn{1}{c|}{0.578±0.047} &\multicolumn{1}{c|}{0.738±0.029}&   \multicolumn{1}{c|}{0.567±0.048} &\multicolumn{1}{c|}{0.239±0.096}
			&\multicolumn{1}{c}{0.652±0.045}\\
			
			\multicolumn{1}{l||}{HAR-Net~\cite{gao2020reconstruction}} &  \multicolumn{1}{c|}{0.793±0.033} &  
			\multicolumn{1}{c|}{0.576±0.047} &\multicolumn{1}{c|}{0.738±0.029}&   \multicolumn{1}{c|}{0.569±0.047} &\multicolumn{1}{c|}{0.234±0.091}
			&\multicolumn{1}{c}{0.653±0.042}\\
			
			\hline
			
			\multicolumn{1}{l||}{SASR (Our Method)} &  \multicolumn{1}{c|}{$\textbf{0.807±0.036}$}& 
	\multicolumn{1}{c|}{$\textbf{0.649±0.049}$} &\multicolumn{1}{c|}{$\textbf{0.775±0.030}$}&   \multicolumn{1}{c|}{$\textbf{0.594±0.060}$} &\multicolumn{1}{c|}{0.269±0.093}  &\multicolumn{1}{c}{$\textbf{0.684±0.055}$}\\
			\hline
			
	\end{tabular}}
	\label{table3}
\end{table*}

\begin{table*}[t]
	\centering
	\caption{\textbf{Vessel} segmentation performance between the reconstructed HR images and the reference HR images on the \textbf{SURE-Z} dataset.}
	\setlength{\tabcolsep}{2.5mm}{
		\begin{tabular}{lllllllll} 
			\hline
			\multicolumn{1}{l||}{} & \multicolumn{6}{c}{\textbf{SURE-Z}}\\
			\hline
			\multicolumn{1}{l||}{Methods}& \multicolumn{1}{c|}{\textit{AUC}  $\uparrow$ }&
			\multicolumn{1}{c|}{\textit{Sen} $\uparrow$}&\multicolumn{1}{c|}{\textit{G-mean} $\uparrow$} &\multicolumn{1}{c|}{\textit{Kappa} $\uparrow$}
			&\multicolumn{1}{c|}{\textit{FDR} $\downarrow$}  &  
			\multicolumn{1}{c}{\textit{Dice} $\uparrow$} \\
			\hline 
			\multicolumn{1}{l||}{HR} &  \multicolumn{1}{c|}{0.941±0.015}&   
			 \multicolumn{1}{c|}{0.796±0.044} &\multicolumn{1}{c|}{0.877±0.024}&  \multicolumn{1}{c|}{0.793±0.035}&  \multicolumn{1}{c|}{0.081±0.025}&  \multicolumn{1}{c}{0.852±0.028} \\
			 \hline
			
			\multicolumn{1}{l||}{Bicubic~\cite{keys1981cubic}} &  \multicolumn{1}{c|}{0.730±0.040}&  
				\multicolumn{1}{c|}{0.509±0.037} &\multicolumn{1}{c|}{0.675±0.028}
			&   \multicolumn{1}{c|}{0.434±0.050} &\multicolumn{1}{c|}{$\textbf{0.306±0.059}$}&\multicolumn{1}{c}{0.586±0.040}\\
			
			\multicolumn{1}{l||}{ScSR~\cite{yang2010image}} &  \multicolumn{1}{c|}{0.733±0.039}&  
			\multicolumn{1}{c|}{0.525±0.040} &\multicolumn{1}{c|}{0.681±0.027}
			&   \multicolumn{1}{c|}{0.436±0.048} &\multicolumn{1}{c|}{0.319±0.059}&\multicolumn{1}{c}{0.592±0.041}\\
			
			\multicolumn{1}{l||}{SRCNN~\cite{dong2015image}} &  \multicolumn{1}{c|}{0.731±0.038}&  
		\multicolumn{1}{c|}{0.531±0.044} &\multicolumn{1}{c|}{0.683±0.028}
			&\multicolumn{1}{c|}{0.434±0.047} &\multicolumn{1}{c|}{0.326±0.056}
			&\multicolumn{1}{c}{0.593±0.041}\\

			\multicolumn{1}{l||}{EDSR~\cite{lim2017enhanced}} &  \multicolumn{1}{c|}{0.733±0.040}&    
		\multicolumn{1}{c|}{0.532±0.041} &\multicolumn{1}{c|}{0.683±0.027}
			&   \multicolumn{1}{c|}{0.433±0.049} &\multicolumn{1}{c|}{0.328±0.060}&\multicolumn{1}{c}{0.593±0.041} \\
			
			\multicolumn{1}{l||}{DRLN~\cite{anwar2020densely}} &  \multicolumn{1}{c|}{0.733±0.040}&    
		\multicolumn{1}{c|}{0.528±0.040} &\multicolumn{1}{c|}{0.682±0.028}&   \multicolumn{1}{c|}{0.433±0.049} &\multicolumn{1}{c|}{0.325±0.061}
			&\multicolumn{1}{c}{0.591±0.042}\\
					\multicolumn{1}{l||}{ RDN~\cite{zhang2018residual} (baseline)} &  \multicolumn{1}{c|}{0.734±0.040}&    
			\multicolumn{1}{c|}{0.526±0.040} &\multicolumn{1}{c|}{0.681±0.027}
			&   \multicolumn{1}{c|}{0.433±0.048} &\multicolumn{1}{c|}{0.322±0.060}&\multicolumn{1}{c}{0.591±0.040} \\
				\hline
				
				\multicolumn{1}{l||}{ZSSR~\cite{shocher2018zero}} &  \multicolumn{1}{c|}{0.732±0.039}&   
				\multicolumn{1}{c|}{0.522±0.038} & 
				\multicolumn{1}{c|}{0.680±0.026}& 
			\multicolumn{1}{c|}{0.435±0.047}  & 
			\multicolumn{1}{c|}{0.317±0.058} &
			\multicolumn{1}{c}{0.590±0.039}\\
		
			\multicolumn{1}{l||}{RealSR~\cite{ji2020real}} & 
			\multicolumn{1}{c|}{0.734±0.042}&    
			\multicolumn{1}{c|}{0.515±0.041}&   \multicolumn{1}{c|}{0.678±0.028} 
			&\multicolumn{1}{c|}{0.438±0.049}&\multicolumn{1}{c|}{0.306±0.056}& 
			\multicolumn{1}{c}{0.590±0.041}\\
			
				\multicolumn{1}{l||}{CycleGAN~\cite{zhu2017unpaired}} &  \multicolumn{1}{c|}{0.725±0.043}&    
	    	\multicolumn{1}{c|}{0.536±0.064} &\multicolumn{1}{c|}{0.682±0.038}&   \multicolumn{1}{c|}{0.428±0.056} &\multicolumn{1}{c|}{0.337±0.055}
			&\multicolumn{1}{c}{0.591±0.055}\\
			
				\multicolumn{1}{l||}{Two-stage GAN~\cite{zhou2019image}} &  \multicolumn{1}{c|}{0.732±0.037}&    
	    	\multicolumn{1}{c|}{0.522±0.039} &\multicolumn{1}{c|}{0.680±0.027}&   \multicolumn{1}{c|}{0.436±0.048} &\multicolumn{1}{c|}{0.316±0.059}
			&\multicolumn{1}{c}{0.591±0.040}\\
			
			\multicolumn{1}{l||}{HAR-Net~\cite{gao2020reconstruction}} &  \multicolumn{1}{c|}{0.734±0.041} &  
			\multicolumn{1}{c|}{0.527±0.040} &\multicolumn{1}{c|}{0.680±0.028}&   \multicolumn{1}{c|}{0.432±0.049} &\multicolumn{1}{c|}{0.325±0.062}
			&\multicolumn{1}{c}{0.590±0.041}\\
			
			\hline
			
			\multicolumn{1}{l||}{SASR (Our Method)} &  \multicolumn{1}{c|}{$\textbf{0.747±0.038}$}& 
			\multicolumn{1}{c|}{$\textbf{0.541±0.049}$} &\multicolumn{1}{c|}{$\textbf{0.688±0.030}$}&   \multicolumn{1}{c|}{$\textbf{0.442±0.051}$} &\multicolumn{1}{c|}{0.323±0.051}  &\multicolumn{1}{c}{$\textbf{0.601±0.045}$}\\
			\hline
			
	\end{tabular}}
	\label{table4}
\end{table*}

\begin{figure*}[t]
\centering{
\includegraphics[width=18cm]{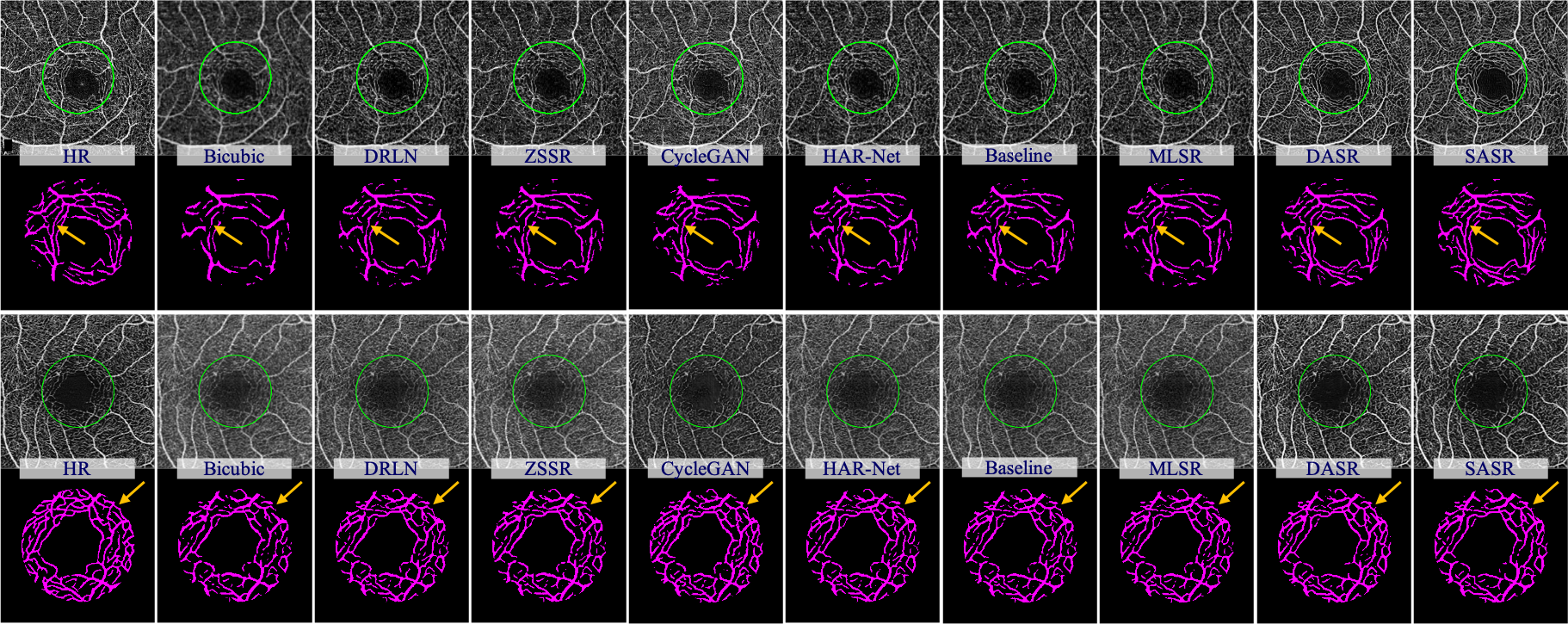}
}
\caption{Vessel segmentation performance on peri-macular region (green circle) of reconstructed OCTA image by different methods over two sample OCTA images selected from SURE-O and SURE-Z sets.}
\label{fig.9}
\end{figure*}

\subsection{Evaluation Metrics}
For the pairs of realistic OCTA images, it is unreliable to evaluate the image quality improvement with the common quality measures, due to that the inherent variations of signal and noise levels can be generated during the image acquisition at different scales, i.e. $3\times3~mm^2$ and $6\times6~mm^2$.
Nevertheless, the similarity on clinically significant structures such as blood vessels provides us with extra possibilities for quality evaluation, i.e., by considering the improvements to FAZ and vascular segmentations. 

For the synthetic OCTA images, we use peak signal-to-noise ratio (PSNR)~\cite{hore2010image} and structural similarity (SSIM)~\cite{wang2004image} and learned perceptual image patch similarity (LPIPS)~\cite{zhang2018unreasonable} to evaluate the reconstruction quality of different super-resolution methods. The PSNR is defined as by:
\begin{equation} 
\operatorname{PSNR}(f, g)=10 \log _{10}\left(255^{2} / M S E(f, g)\right)
\end{equation}
\begin{equation} 
\operatorname{MSE}(f, g)=\frac{1}{M N} \sum_{i=1}^{M} \sum_{j=1}^{N}\left(f_{i j}-g_{i j}\right)^{2}
\end{equation}
where $f$ and $g$ present reference and reconstructed images, both of size $M \times N$.  The SSIM is defined as: 

\begin{equation} 
\operatorname{SSIM}(f, g)=\frac{\left(2 \mu_{f} \mu_{g}+c_{1}\right)\left(2 \sigma_{f g}+c_{2}\right)}{\left(\mu_{f}^{2}+\mu_{g}^{2}+c_{1}\right)\left(\sigma_{f}^{2}+\sigma_{g}^{2}+c_{2}\right)}
\end{equation}

where $\mu_{f}$ and $ \mu_{g}$ are the mean values of $f$ and $g$, respectively. $\sigma_{f}$ and $\sigma_{g}$ denote the variance of $f$ and $g$, respectively.  $\sigma_{f g}$ represents the covariance of $f$ and $g$. $c_{1}$ and  $c_{2}$ are the constants that maintain stability. PSNR and SSIM are commonly used metrics for the evaluation of image super-resolution, and they pay attention to image fidelity rather than visual quality. In contrast, LPIPS is more concerned with whether the visual features of the images are similar. LPIPS uses pre-trained AlexNet to extract image features and then calculates the distance between two features. Therefore, the smaller the LPIPS, the closer the generated image is to the ground truth. 

\subsection{Performances on realistic low resolution image}
\label{subsec:EvaluationRealistic}
In this subsection, we evaluated the reconstruction performance when the realistic low resolution image was utilized as input. 
To prove the superiority of the proposed method, the following methods were compared:

(1) \textbf{Traditional methods for natural synthetic images}: the Bicubic method~\cite{keys1981cubic} and Sparse coding based Super Resolution (ScSR)~\cite{yang2010image}.

(2) \textbf{Deep learning methods for natural synthetic images}: Super-Resolution Convolutional Neural Network (SRCNN)~\cite{dong2015image}, Enhanced Deep Super-Resolution network (EDSR)~\cite{lim2017enhanced}, Densely Residual Laplacian Network (DRLN)~\cite{anwar2020densely} and Residual Dense Network (RDN)~\cite{zhang2018residual}.

(3) \textbf{Deep learning methods for realistic images}: Zero-Shot Super-Resolution (ZSSR)~\cite{shocher2018zero}, Realistic degradation framework for Super-Resolution (RealSR)~\cite{ji2020real}, CycleGAN~\cite{zhu2017unpaired}, Two-stage GAN~\cite{zhou2019image} and High-resolution Angiogram Reconstruction Network (HAR-Net)~\cite{gao2020reconstruction}. Particularly, the HAR-Net was specialized in OCTA image super-resolution task.
\subsubsection{Performance of reconstruction results from realistic OCTA images}

Fig.~\ref{fig.7} shows the visual effect of the super-resolution results, and our method shows relatively better performance in terms of feature preservation and contrast. This might be because our method pays more attention to multi-scale structure and edge details. 
For a careful observation, our method yields better vascular details compared to other comparative methods. The main reason is that the proposed network emphasizes the vascular signal ignoring the effect of background noise during the reconstruction process. 
As can be seen from the Fig.~\ref{fig.7}, the reconstruction results of our method avoid the generation of redundant vascular information. The vague capillary details in LR images are properly reconstructed in the generated HR images with better visibility. We also reported the number of parameters and test times of single image for different super-resolution models, as shown in Table~\ref{paras}. Compared to the RDN, the time difference of our method between the tests on the two sets is small, despite the increased parameters of the proposed model. This is because the discriminative network is not involved in the testing. The testing time of the proposed model is within allowable limits in contrast to other adversarial migration methods.

\subsubsection{Performance of FAZ segmentation}
The FAZ is a highly specialized region and useful marker that indicates fovea health. To confirm the reconstruction performance on realistic images, we further performed FAZ segmentation of the reconstructed images on two OCTA sets.  
Specifically, we employed a trained OCTA-Net~\cite{ma2020rose} to segment the FAZ in both of the reconstructed HR images and the reference HR images. This model had been pre-trained on HR images with the ground truth of FAZ contours. Note that the ground truth was obtained by manual segmentation in the reference HR image, and used to calculate the FAZ segmentation metrics for the reconstructed and reference images.

\begin{table}[t]
\centering
\caption{Comparison of parameter numbers and test times for different super-resolution models on two sets.}
\begin{tabular}{llll}
\hline
\textbf{Method} & \textbf{Parameter} & \textbf{Time (SURE-Z)} & \textbf{Time (SURE-O)} \\ \hline
SRCNN           & 57K                         & 0.021s                      & 0.017s                      \\
RDN             & 5.5M                        & 0.257s                      & 0.224s                      \\
EDSR            & 1.37M                       & 0.049s                      & 0.045s                      \\
DRLN            & 34.43M                      & 0.584s                      & 0.529s                      \\
ZSSR            & 225K                        & 0.015s                      & 0.014s                      \\
CycleGAN        & 89.17M                      & 1.661s                       & 0.561s                      \\
HAR-Net         & 851K                        & 0.043s                      & 0.041s                      \\
Ours            & 13M                         & 0.319s                      & 0.278s                      \\ \hline
\end{tabular}
\label{paras}
\end{table}

\begin{table}[t]
\centering
\caption{Comparison of \textit{VLD} and \textit{VT} in reconstructed images from different methods on two OCTA sets.}
\setlength{\tabcolsep}{3mm}{
\begin{tabular}{llllll} 
\hline
\multicolumn{1}{l||}{} & \multicolumn{2}{c||}{\textbf{SURE-O}} & \multicolumn{2}{c}{\textbf{SURE-Z}}\\
\hline
\multicolumn{1}{l||}{Methods} & \multicolumn{1}{c|}{\textit{VT}}&\multicolumn{1}{c||
}{\textit{VLD}} &\multicolumn{1}{c|}{\textit{VT}} & \multicolumn{1}{c}{\textit{VLD}} \\
\hline 
\multicolumn{1}{l||}{Ground Truth} &  \multicolumn{1}{c|}{1.599}&  \multicolumn{1}{c||}{10.857} &  \multicolumn{1}{c|}{2.389}&  \multicolumn{1}{c}{8.336}\\

\multicolumn{1}{l||}{HR} &  \multicolumn{1}{c|}{1.578}&  \multicolumn{1}{c||}{7.909} &   \multicolumn{1}{c|}{2.175}&  \multicolumn{1}{c}{7.823} \\
\hline
\multicolumn{1}{l||}{Bicubic} &  \multicolumn{1}{c|}{1.962}&  \multicolumn{1}{c||}{5.209} &  \multicolumn{1}{c|}{2.098}&  \multicolumn{1}{c}{6.546} \\

\multicolumn{1}{l||}{ScSR} &  \multicolumn{1}{c|}{2.010}&  \multicolumn{1}{c||}{6.502} &   \multicolumn{1}{c|}{2.096}&  \multicolumn{1}{c}{7.214} \\

\multicolumn{1}{l||}{SRCNN} &  \multicolumn{1}{c|}{2.095}&  \multicolumn{1}{c||}{6.777} &   \multicolumn{1}{c|}{2.043}&  \multicolumn{1}{c}{7.624} \\

\multicolumn{1}{l||}{RDN} &  \multicolumn{1}{c|}{1.939}&  \multicolumn{1}{c||}{6.891} &  \multicolumn{1}{c|}{2.040}&  \multicolumn{1}{c}{7.521}  \\

\multicolumn{1}{l||}{EDSR} &  \multicolumn{1}{c|}{1.965}&  \multicolumn{1}{c||}{6.829} &  \multicolumn{1}{c|}{2.072}&  \multicolumn{1}{c}{7.752}  \\

\multicolumn{1}{l||}{DRLN} &  \multicolumn{1}{c|}{1.852}&  \multicolumn{1}{c||}{6.822} &  \multicolumn{1}{c|}{2.077}&  \multicolumn{1}{c}{7.592} \\
\hline 

\multicolumn{1}{l||}{ZSSR} &  \multicolumn{1}{c|}{1.953}&  \multicolumn{1}{c||}{6.862}&  \multicolumn{1}{c|}{2.052}&  \multicolumn{1}{c}{7.153} \\

\multicolumn{1}{l||}{RealSR} &  \multicolumn{1}{c|}{2.065}&  \multicolumn{1}{c||}{6.703}&  \multicolumn{1}{c|}{2.076}&  \multicolumn{1}{c}{6.814}\\

\multicolumn{1}{l||}{CycleGAN} &  \multicolumn{1}{c|}{2.085}&  \multicolumn{1}{c||}{7.112}&  \multicolumn{1}{c|}{2.083}&  \multicolumn{1}{c}{7.543}\\

\multicolumn{1}{l||}{Two-stage GAN} &  \multicolumn{1}{c|}{1.869}&  \multicolumn{1}{c||}{6.974}&  \multicolumn{1}{c|}{2.073}&  \multicolumn{1}{c}{7.173}\\

\multicolumn{1}{l||}{HAR-Net} &  \multicolumn{1}{c|}{2.106}&  \multicolumn{1}{c||}{6.740} &  \multicolumn{1}{c|}{2.097}&  \multicolumn{1}{c}{7.574} \\

\hline
\multicolumn{1}{l||}{Ours} &  \multicolumn{1}{c|}{1.724}&  \multicolumn{1}{c||}{8.695}  &  \multicolumn{1}{c|}{2.123}&  \multicolumn{1}{c}{8.140}  \\
\hline

\end{tabular}}
\label{VLD}
\end{table}
The following metrics are calculated and compared:\\
$\bullet$ Area Under the ROC Curve (AUC); \\
$\bullet$ Sensitivity (SEN) = TP / (TP + FN);\\
$\bullet$ Accuracy (ACC) = (TP + TN) / (TP + TN + FP + FN); \\
$\bullet$  $\textit{Kappa}$ score = $(Accuracy - p_e) / (1 - p_e)$;  \\ 
$\bullet$ False Discovery Rate (FDR) = FP / (FP + TP); \\
$\bullet$ $\textit{G-mean}$ score = $\rm \sqrt{Sensitivity \times Specificity}$; \\  
$\bullet$ Dice coefficient (Dice) = 2 $\times$ TP / (FP + FN + 2 $\times$ TP);\\
where TP is true positive, FP is false positive, TN is true negative, and FN is false negative.
$p_e$ in $\textit{Kappa}$ score represents opportunity consistency between the ground truth and prediction, and it is denoted as:
\begin{equation} 
\begin{aligned} 
p_e = & ((TP + FN)(TP + FP) + (TN + FP)(TN + FN)) \\ & / (TP + TN + FP + FN)^2
\end{aligned}
\end{equation}
$\bullet$ Hausdorff distance ($d_\text{HD}$)~\cite{karimi2019reducing} is denoted as :
\begin{equation} 
\begin{aligned} 
\operatorname{d_\text{HD}}(X, Y)=\max _{x \in X} \min _{y \in Y}\|x-y\|_{2}
\end{aligned}
\end{equation}
where $d_\text{HD}$ calculates the directed Hausdorff distance between the point set $Y$ of edge contour in the reconstructed or reference HR image and the point set $X$ of edge contour in the ground truth.


\begin{table*}[t]\footnotesize
 \centering \caption{Ablation studies of FAZ segmentation on both the \textbf{SURE-O} and \textbf{SURE-Z} subsets}
 \setlength{\tabcolsep}{1.5mm}{ 
 \begin{tabular}{l||c|c|c|c|c|c|c||c|c|c|c|c|c|c} 
 \hline  
 & \multicolumn{7}{c||}{\textbf{SURE-O}} & \multicolumn{7}{c}{\textbf{SURE-Z}}\\  
  \hline  
  Methods &  \textit{AUC}$\uparrow$  & \textit{ACC}$\uparrow$  & \textit{G-mean}$\uparrow$  & \textit{Kappa}$\uparrow$ &\textit{FDR}$\downarrow$ &\textit{Dice}$\uparrow$ &\textit{$d_\text{HD}$}$\downarrow$ &\textit{AUC}$\uparrow$ &\textit{ACC} $\uparrow$ & \textit{G-mean}$\uparrow$ & \textit{Kappa}$\uparrow$  &\textit{FDR}$\downarrow$ & \textit{Dice}$\uparrow$ &\textit{$d_\text{HD}$}$\downarrow$\\  
  \hline 
 RDN~\cite{zhang2018residual} (baseline) &  0.942 &  0.846 &0.913 & 0.863 & 0.110 & 0.867 & 29.12 &  0.941 & 0.830 & 0.908 & 0.840 & 0.134 & 0.846 & 69.92 \\ 
  MLSR & 0.945 &  0.854 & 0.918 & 0.867 & 0.108 &  0.871 & 31.01 & 0.943 & 0.834 & 0.910 & 0.845 & 0.129 & 0.850 & 76.97 \\ 
 DASR &  0.955 &  0.880 & 0.935 & $\textbf{0.903}$&  $\textbf{0.066}$ & $\textbf{0.906}$ & 23.57 & 0.953 & 0.877 & 0.934 & 0.893 & $\textbf{0.081}$ & 0.897 & 49.95 \\ 
 \hline 
SASR (Our Method) & $\textbf{0.961}$ & $\textbf{0.890}$ & $\textbf{0.940}$ & 0.886 & 0.104 & 0.890 & $\textbf{17.58}$ & $\textbf{0.958}$& $\textbf{0.889}$ & $\textbf{0.940}$ & $\textbf{0.897}$ & 0.086 & $\textbf{0.900}$ & $\textbf{41.01}$ \\  
\hline 
\end{tabular}}   
\label{ablation-faz}   
\vspace{-10pt}
\end{table*}

\begin{table*}[t]\footnotesize
 \centering \caption{Ablation studies of vessel segmentation on both the \textbf{SURE-O} and \textbf{SURE-Z} subsets}
 \setlength{\tabcolsep}{1.5mm}{ 
 \begin{tabular}{l||c|c|c|c|c|c||c|c|c|c|c|c} 
 \hline  
 & \multicolumn{6}{c||}{\textbf{SURE-O}} & \multicolumn{6}{c}{\textbf{SURE-Z}}\\  
  \hline  
  Methods &  \textit{AUC}$\uparrow$ & \textit{ACC}$\uparrow$ & \textit{G-mean}$\uparrow$ & \textit{Kappa}$\uparrow$ &\textit{FDR}$\downarrow$ &\textit{Dice}$\uparrow$ &\textit{AUC}$\uparrow$ &\textit{ACC}$\uparrow$ & \textit{G-mean}$\uparrow$ & \textit{Kappa}$\uparrow$  &\textit{FDR}$\downarrow$ & \textit{Dice}$\uparrow$\\  
  \hline 
 RDN~\cite{zhang2018residual} (baseline) &  0.793 &  0.740 & 0.571 & 0.740 & $\textbf{0.235}$ & 0.655 & 0.734 & 0.526 &  0.681 & 0.433 & $\textbf{0.322}$  & 0.591 \\ 
  MLSR & 0.793 &  0.616 & 0.759 & 0.582 & 0.256 &  0.671 & 0.735 & 0.529 & 0.681 & 0.432 & 0.326 & 0.591  \\ 
 DASR &  0.800 &  0.624 & 0.762 & 0.585 & 0.260 & 0.673 & 0.740 & $\textbf{0.544}$ & 0.687 & 0.433 & 0.336 & 0.597 \\ 
 \hline 
SASR (Our Method) & $\textbf{0.807}$ & $\textbf{0.649}$ & $\textbf{0.775}$ & $\textbf{0.594}$ & 0.269 & $\textbf{0.684}$ & $\textbf{0.747}$ & 0.541 & $\textbf{0.688}$ & $\textbf{0.442}$ & 0.323 & $\textbf{0.601}$ \\  
\hline 
\end{tabular}}   
\label{ablation-vessel}   
\vspace{-10pt}
\end{table*}

\begin{table}[t]
\centering
\caption{The quantitative results of super-resolution reconstruction using different approaches on two synthetic OCTA sets.}
\setlength{\tabcolsep}{0.5mm}{
\begin{tabular}{lllllllll} 
\hline
\multicolumn{1}{l||}{} & \multicolumn{3}{c||}{\textbf{SURE-O}} & \multicolumn{3}{c}{\textbf{SURE-Z}}\\
\hline
\multicolumn{1}{l||}{Methods}  &  \multicolumn{1}{c|}{\textit{PSNR}  $\uparrow$ }&\multicolumn{1}{c|}{\textit{SSIM $\uparrow$}} & \multicolumn{1}{c||}{\textit{LPIPS} $\downarrow$} &\multicolumn{1}{c|}{\textit{PSNR} $\uparrow$} & \multicolumn{1}{c|}{\textit{SSIM} $\uparrow$} & \multicolumn{1}{c}{\textit{LPIPS} $\downarrow$} \\
\hline 
\multicolumn{1}{l||}{Bicubic} &  \multicolumn{1}{c|}{15.09}&  \multicolumn{1}{c|}{0.634} &  \multicolumn{1}{c||}{0.304} &  \multicolumn{1}{c|}{20.57}&  \multicolumn{1}{c|}{0.813} & \multicolumn{1}{c}{0.569} \\

\multicolumn{1}{l||}{SRCNN} &  \multicolumn{1}{c|}{19.46}&  \multicolumn{1}{c|}{0.745} &  \multicolumn{1}{c||}{0.236} &  \multicolumn{1}{c|}{27.27}&  \multicolumn{1}{c|}{0.926} & \multicolumn{1}{c}{0.075}  \\

\multicolumn{1}{l||}{RDN} &  \multicolumn{1}{c|}{19.97}&  \multicolumn{1}{c|}{0.770} &  \multicolumn{1}{c||}{0.179} &  \multicolumn{1}{c|}{32.12}&  \multicolumn{1}{c|}{0.976} & \multicolumn{1}{c}{0.022}  \\

\multicolumn{1}{l||}{EDSR} &  \multicolumn{1}{c|}{19.93}&  \multicolumn{1}{c|}{0.770} &  \multicolumn{1}{c||}{0.185} &  \multicolumn{1}{c|}{30.92}&  \multicolumn{1}{c|}{0.968} & \multicolumn{1}{c}{0.042}  \\

\multicolumn{1}{l||}{DRLN} &  \multicolumn{1}{c|}{20.08}&  \multicolumn{1}{c|}{$\textbf{0.773}$} &  \multicolumn{1}{c||}{0.169} &  \multicolumn{1}{c|}{31.19}&  \multicolumn{1}{c|}{0.978} & \multicolumn{1}{c}{0.023} \\
\hline 

\multicolumn{1}{l||}{Two-stage GAN} &  \multicolumn{1}{c|}{18.67}&  \multicolumn{1}{c|}{0.706} &  \multicolumn{1}{c||}{0.164} &  \multicolumn{1}{c|}{29.96}&  \multicolumn{1}{c|}{0.916} & \multicolumn{1}{c}{0.067} \\

\multicolumn{1}{l||}{HAR-Net} &  \multicolumn{1}{c|}{19.95}&  \multicolumn{1}{c|}{0.767} &  \multicolumn{1}{c||}{0.187} &  \multicolumn{1}{c|}{30.43}&  \multicolumn{1}{c|}{0.964} & \multicolumn{1}{c}{0.038} \\

\hline
\multicolumn{1}{l||}{ MLSR } &  \multicolumn{1}{c|}{$\textbf{20.34}$}&  \multicolumn{1}{c|}{0.770} &  \multicolumn{1}{c||}{$\textbf{0.142}$} &  \multicolumn{1}{c|}{$\textbf{34.28}$}&  \multicolumn{1}{c|}{$\textbf{0.988}$} & \multicolumn{1}{c}{$\textbf{0.008}$}  \\
\hline

\end{tabular}}
\label{table5}
\end{table}

Fig.~\ref{fig.8} shows the FAZ segmentation results of the reconstructed images using different methods on the two OCTA sets, where the red area represents the under-segmentation results while the blue area represents the over-segmentation results.
It can be seen that the FAZ segmentation results of our method are closest to the HR segmentation results and the ground truth, where the under-segmented and over-segmented areas are the smallest.
In addition, Table~\ref{table1} and \ref{table2} provide the FAZ segmentation results of all methods on the realistic images of the two sets, respectively. 
In both sets, our method is far ahead of the other methods in terms of all metrics (results in boldface).
Compared with the baseline network RDN, the segmentation metric AUC improves by 1.9\%  and 1.7\%, G-mean improves by 2.7\% and 3.2\%, Kappa improves by 2.3\%  and 5.7\%, and $d_\text{HD}$ improves by 11.54 and 28.91 on two sets, respectively. These improvements fully demonstrate the superiority of the proposed modules in our network. 
We also compared our method with HAR-Net, which specializes in OCTA image super-resolution reconstruction task. The evaluation results show that our proposed network achieves better performance in terms of all metrics. Meanwhile, SEN improves by 4.5\% and 6.0\%, FDR improves by 0.6\%  and 4.8\%, and Dice improves by 2.4\%  and 5.5\% on both sets, respectively.

\subsubsection{Performance of capillary segmentation}
Capillary segmentation is considered as another important evaluation method for the super-resolution reconstruction on realistic OCTA images. Thus, we compared the segmentation results based on the reconstructions from all different methods.
Similarly, we employed the trained vessel segmentation model (OCTA-Net) to extract the small capillaries in both of the reconstructed HR images and the reference HR images. This model had been pre-trained on HR images with manually labeled vascular networks. 
To evaluate the performance of segmentation after reconstruction, we used the following metrics similar to FAZ segmentation: AUC, SEN, G-mean, Kappa, FDR and Dice. In addition, we calculated vascular-related measures to verify the consistency of the super-resolution results from different perspectives: 
1) Vascular Length Density (VLD): the ratio between the total number of microvascular centreline pixels and the area of analyzed region.
2) Vascular Tortuosity (VT): a metric to measure the tortuous level of vascularture, computed by applying the method proposed by~\cite{zhao2020automated}.

It is important to emphasize that the region for segmentation comparison is defined within the $1.25\times1.25~mm^2$ FOV around the macula for the following reasons:
(1) The area around the macula is more correlated with disease;
(2) The small capillaries in this selected region have better visibility and can produce a higher segmentation confidence level for evaluation purpose.

The top row and the third row of Fig.~\ref{fig.9} show reconstruction results using different methods on two sets.
It can been observed that more capillaries are identified in our SASR reconstructed images, where the improvements are indicated by yellow arrows. The vessel segmentation results are also more accurate due to the significantly improved image quality between the vessels and the background area of the reconstructed images.
From the segmentation results in Table~\ref{table3} and Table~\ref{table4}, we can observe that our approach achieves the best performance on both sets. Specifically, Our method outperforms the state-of-the-art DRLN by 1.3\% and 1.4\% in AUC, 7.0\% and 1.3\% in SEN, and 2.9\% and 1.0\% in Dice, respectively.
Compared with ZSSR, SEN is 7.5\% and 1.9\% higher, G-mean is 3.8\% and 0.8\% higher, Dice is 3.2\% and 1.1\% higher, respectively.
Moreover, compared with CycleGAN, our method outperforms it in all metrics on both sets. This is because unpaired transfer learning lacks the constraint of structural consistency, while the proposed sparse edge-aware loss enhances the texture details of the reconstructed images.
Our method also outperforms the baseline model by 7.0\% and 1.5\% in SEN, 3.5\% and 0.7\% in G-mean, and 2.3\% and 0.9\% in Kappa, respectively. These quantitative evaluations prove that our module has great improvement compared to the baseline network. These improvements in performance are consistent with the segmentation results indicated by the yellow arrows in Fig.~\ref{fig.9}, where the segmentation model can successfully extract small capillaries with good continuity and integrity from the reconstructed images of the proposed method, while the reconstructed images of the other methods yield relatively low capillary correspondence. Furthermore, Table~\ref{VLD} reports the vascular measurement metric results for the different methods. The results show that the VT and VLD results of the proposed method  are closest to the high-resolution images on both sets. 



\subsection{Performances on synthetic low resolution image}
In this section, we evaluated the reconstruction performance when the synthetic low resolution image was utilized as input. 

Table~\ref{table5} shows the super-resolution reconstruction results of all methods on the two synthetic sets. The experiments show that our method achieves the best performance in terms of PSNR on both sets, which represents that the fidelity of our method outperforms other methods. The LPIPS metric of our method also presents the highest performance on two OCTA dataset, indicating our results are much closer to the HR images in terms of visual characteristics. 
We adopted a simple method for image degradation, but the reconstruction effect of SURE-O dataset is far inferior to that of SURE-Z data, which is mainly because the  image quality of SURE-Z dataset is relatively higher, and has less background noise.
For SURE-Z dataset, all methods achieve comparable performance in terms of their reconstruction results. Nevertheless, our proposed network still achieves the best results in all metrics. 
When compared with EDSR, the proposed method outperforms it in all metrics on both sets, with a respectively higher PSNR of 0.41 dB and 3.36 dB. The LPIPS is also improved by 4.3\% and 3.4\%, respectively.
In addition, compared with the HAR-Net, our method achieves significant improvements in all three metrics, with 0.39 dB and 3.85 dB higher on PSNR,  0.3\% and 2.4\% higher on SSIM, 4.5\% and 3.0\% higher on LPIPS, respectively. 

\subsection{Ablation studies}
In this paper, our proposed method employed three modules to establish the super-resolution framework, i.e., multi-level super-resolution model (MLSR), patch discrimination, and a new sparse edge-aware loss. 
To evaluate the effectiveness of each module, we validated the  reconstruction performance using different combinations of these modules on two OCTA sets separately.

\subsubsection{Multi-level super-resolution model (MLSR)} The effectiveness of the MLSR module was demonstrated by comparing with different reconstruction results separately on the synthetic and realistic OCTA sets. 
For the synthetic data in both sets, Table~\ref{table5} shows that MLSR outperforms RDN in all metrics on both sets, with a respectively higher PSNR of 0.37 dB and 2.16 dB. The LPIPS is also improved by 3.7\% and 1.4\%, respectively.
For the realistic data, Table~\ref{table1}-\ref{table4} show that the MLSR achieves higher performance on both FAZ and vessel segmentations compared to other approaches. The results indirectly prove that the super-resolution performance of the model on synthetic data can affect the performance tested on realistic data until the domain gap between synthetic and realistic data are mitigated. 


\subsubsection{Patch discrimination} Furthermore, we also validated the effectiveness of patch discrimination in the domain adaptation super-resolution framework. We reported the FAZ and vessel segmentation results on both sets. It is worth while to emphasize that one of the main contributions in this work is the domain adaptation framework, which guides the super-resolution reconstruction of realistic LR images via reducing the domain map between synthetic and realistic data. 
In both OCTA sets, Fig.~\ref{fig.7}-\ref{fig.9} show that the reconstruction performance of DASR outperforms MLSR in terms of the visual effect of super-resolution reconstruction results and the performance of FAZ and vessel segmentation. 
The experimental results in Table~\ref{ablation-faz} and Table~\ref{ablation-vessel} also show that DASR outperforms MLSR in all metrics for the FAZ and vessel segmentation, respectively.
It proves that DASR can enhance the clarity and contrast of the vessels in the reconstructed images, which leads to improvement of the FAZ and vessel segmentation results. 

\subsubsection{Sparse edge-aware loss (SE-loss)} Based on the proposed DASR, we also introduced the SE-loss to further improve the reconstruction results. SE-loss is mainly designed to optimize the reconstruction results by using the highly similar characteristics of blood vessels in LR and HR images. 
Fig.~\ref{fig.9} shows that the proposed SE-loss can alleviate the problem of vascular discontinuity of OCTA reconstruction results and meanwhile enhance the image quality. 
The quantification results of the vessel segmentation in Table~\ref{ablation-vessel} also show that the SE-loss is beneficial to the precise vessel reconstruction in LR images.
In local regions of high similarity between LR and HR images, super-resolution reconstruction is performed by taking advantage of the HR vascular information in a supervised manner. While in regions of low similarity, the reconstruction of LR regions are achieved via unsupervised learning. This can be useful to the enhancement of local visibility with rich capillary details. 

\section{Conclusion}
In general, $3\times3~mm^2$ and $6\times6~mm^2$ FOVs are two typical OCTA acquisition criteria. The scan quality of larger $6\times6~mm^2$ FOV angiography is significantly lower compared to $3\times3~mm^2$ images, which leads to the presence of noise and the invisibility of small capillaries. Such limitations brings challenges to the judgment of the ophthalmologist or researcher when larger FOV is needed. Therefore, we have proposed a super-resolution method to alleviate this problem. 
We attempted to address this problem through a domain adaptation approach and construct two OCTA sets from two different devices. First, we used bicubic method to perform degradation on HR images. We then proposed a super-resolution method based on domain adaptation to reconstruct the realistic OCTA images by reducing the difference between the spatial feature domain of the synthetic and the realistic images.
In our experiments, we evaluated the reconstruction performance in terms of quality improvement and segmentation accuracy separately. 
The experimental results show that 
our method achieves superior performance on both synthetic and realistic images.

In future work, we will extend the proposed single-image super-resolution method to more image modalities such as OCT and ultrasound images for quality improvement. We will also apply our reconstruction technique on clinical data to support dedicated disease analysis, and at the same time improving the reliability and confidence of our method from a clinical perspective.

\bibliographystyle{IEEEtran}
\bibliography{ref}

\end{document}